\documentclass[a4paper,11pt]{article}
\usepackage{jinstpub} 

\usepackage{makecell}
\usepackage[utf8]{inputenc} 
\usepackage[T1]{fontenc}    
\usepackage{hyperref}       
\usepackage{url}            
\usepackage{booktabs}       
\usepackage{amsfonts}       
\usepackage{nicefrac}       
\usepackage{microtype}      
\usepackage{xcolor}         
\usepackage{graphicx}      
\usepackage{xfrac}          
\usepackage{caption}
\usepackage{subcaption}
\usepackage{multirow}

\usepackage{lineno}

\title{Generalizing to new geometries with Geometry-Aware Autoregressive Models (GAAMs) for fast calorimeter simulation}

\author[a]{Junze Liu,}
\author[b,c]{Aishik Ghosh,}
\author[b]{Dylan Smith,}
\author[a]{Pierre Baldi,}
\author[b]{Daniel Whiteson}

\affiliation[a]{Department of Computer Science, University of California, Irvine, CA 92697, USA}
\affiliation[b]{Department of Physics and Astronomy, University of California, Irvine, CA 92697, USA}
\affiliation[c]{Physics Division, Lawrence Berkeley National Laboratory, Berkeley, CA 94720, USA}
\emailAdd{junzel1@uci.edu}

\abstract{Generation of simulated detector response to collision products is crucial to data analysis in particle physics, but computationally very expensive. One subdetector, the calorimeter, dominates the computational time due to the high granularity of its cells and complexity of the interactions. Generative models can provide more rapid sample production, but currently require significant effort to optimize performance for specific detector geometries, often requiring many models to describe the varying cell sizes and arrangements, without the ability to generalize to other geometries. We develop a {\it geometry-aware} autoregressive model, which learns how the calorimeter response varies with geometry, and is capable of generating simulated responses to unseen geometries without additional training. The geometry-aware model outperforms a baseline unaware model by over 50\% in several metrics such as the Wasserstein distance between the generated and the true distributions of key quantities which summarize the simulated response. A single geometry-aware model could replace the hundreds of generative models currently designed for calorimeter simulation by physicists analyzing data collected at the Large Hadron Collider. This proof-of-concept study motivates the design of a foundational model that will be a crucial tool for the study of future detectors, dramatically reducing the large upfront investment usually needed to develop generative calorimeter models.}

\begin{document}
\maketitle
\flushbottom

\section{Introduction}
Collision of particles at high-energy, such as those at the Large Hadron Collider (LHC), provide clues about the fundamental nature of matter and its interactions.  A crucial tool in experimental design and data analysis is the production of vast samples of high-fidelity simulations of the detectors which observe those collisions.  Samples of simulated collisions allow for characterization of the expected response of future detectors as well as simulation-based-inference in data analysis, which often require billions of samples. The highest-fidelity simulations, via Geant4~\cite{GEANT4:2002zbu,Allison:2016lfl,Allison:2006ve}, are produced by detailed time-evolution of individual particles and the microphysics of their interactions with detector elements, which can produce showers of additional particles.  The exponentially increasing number of particles results in large computational expense, limiting the production of  samples of simulated collisions. Efficiency of sample generation will become increasingly vital as the LHC enters an era of high-intensity beams which will produce higher volumes of data. However, the full time evolution of the shower is not observed or recorded, and strategies which directly generate the final cell responses can be significantly cheaper.  Deep generative models (DGM) are a promising approach to automatically learning such a  response function.

Machine learning has been increasingly useful to solve challenging particle physics problems~\cite{baldi2014searching,baldi2021deep,calafiura2022artificial,lu2022resolving,shmakov2023end}. Many experiments, including those at the LHC, have already invested resources in developing DGMs designed to produce shower images for their own calorimeters~\cite{ATLAS:2022jhk,ATLAS:2021pzo,Erdmann:2018jxd,Ratnikov:2020dcm}. However, the calorimeter geometries (i.e., the shape, position, and size of cells and their relative arrangements) not only differ greatly between experiments but can also vary widely from region to region within the same detector. As a result, experiments have to spend significant human resources developing DGM architectures for their specific calorimeter geometries. For example, the ATLAS experiment developed hundreds of generative networks to span different regions of their calorimeter~\cite{ATLAS:2021pzo}. Further, the geometry-specific generative model architectures developed by each experiment vary considerably. While there has recently been significant innovation aimed at improving the accuracy of shower simulations using DGMs for fixed calorimeter geometries ~\cite{Paganini:2017dwg, 3dgan_epj,Buhmann:2020pmy,Krause:2021ilc,Krause:2021wez,Mikuni:2022xry}, the key challenge remains the development of models  capable of generating samples for varying calorimeter geometries, which is essential  to scale these proof-of-concept demonstrations to a deployed product.  This learning task has no analogy in natural images, where pixel sizes are uniform, and  no off-the-shelf machine learning solutions are available. This problem requires a novel approach.

 This paper addresses this crucial computational challenge, presenting  a generative model which learns to interpolate across a training set with
 varying calorimeter geometries and is capable of zero-shot generalization to new geometries on which it has not been trained. This is achieved by training an autoregressive model that is aware of the properties of the calorimeter cell, adjusting the energy deposition based on the size and position of that cell. Such a model could reflect a generalized understanding of calorimeter response, to be shared across experiments with distinct geometries. This would enable individual experiments to fine-tune the model to their specific calorimeter, requiring smaller training samples and  fewer computing resources to  rapidly generate the samples of simulated collisions needed for data analysis. In addition, such a model would allow to rapidly characterize the scientific potential of future detectors. Such general-purpose, pre-trained models are often called `foundational models'.
 

The rest of this paper is organized as follows. Sec.~\ref{sec:relatedWork} discusses related work, Sec.~\ref{sec:Data} describes the dataset used for training and evaluation, Sec.~\ref{sec:genModels} details the ML methods compared, and Sec.~\ref{sec:results} discusses the results, including success with interpolation and failure with extrapolation on novel geometries. The paper ends with a conclusion and future outlook in Sec.~\ref{sec:Conclusion}. 

\begin{figure}
    \centering
    \begin{subfigure}[b]{0.24\textwidth}
        \centering
        \includegraphics[width=\textwidth]{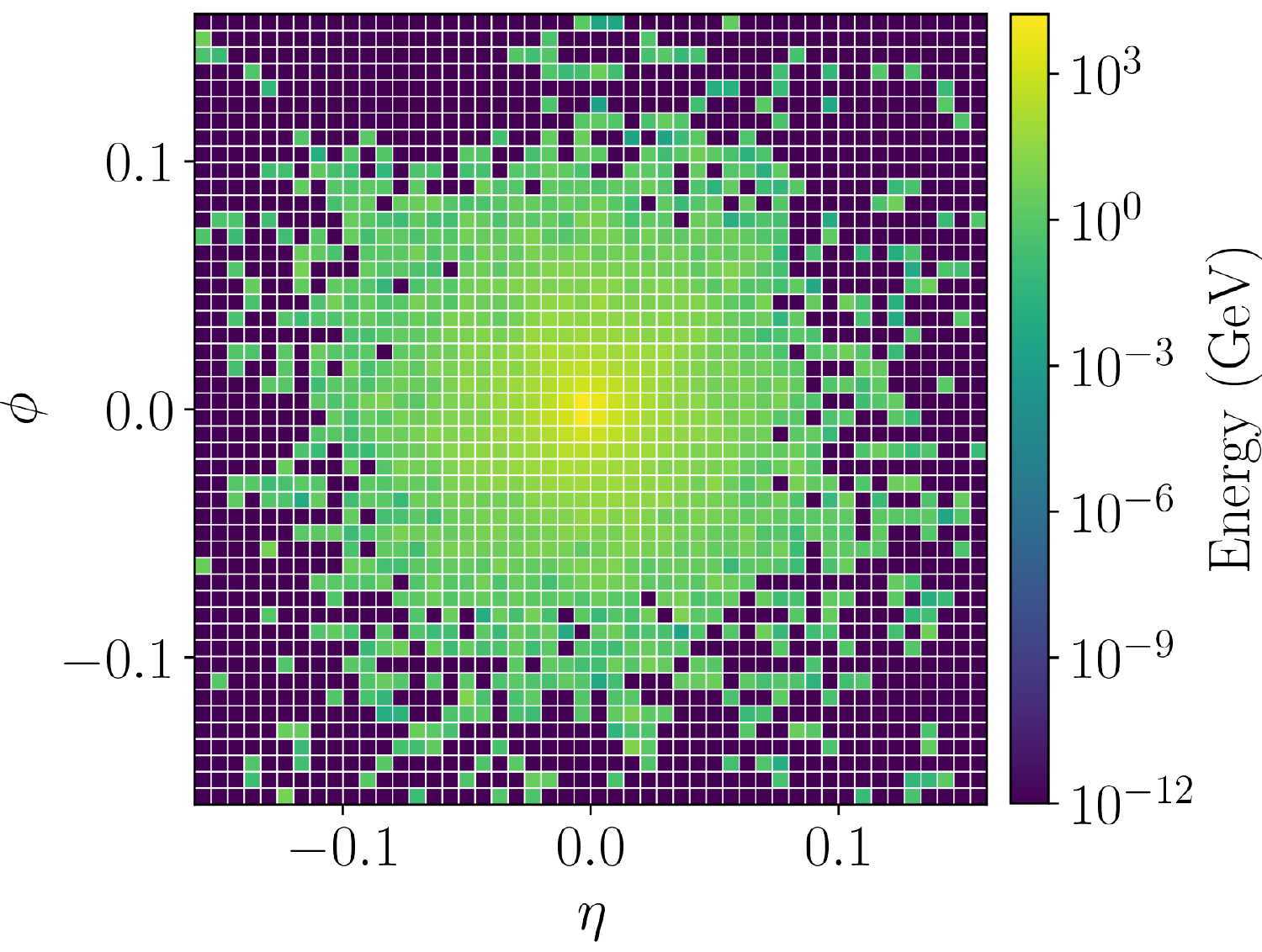}
        \caption{(48, 48)\\ }
        \label{fig:12x12}
    \end{subfigure}
    \hfill
    \begin{subfigure}[b]{0.24\textwidth}
        \centering
        \includegraphics[width=\textwidth]{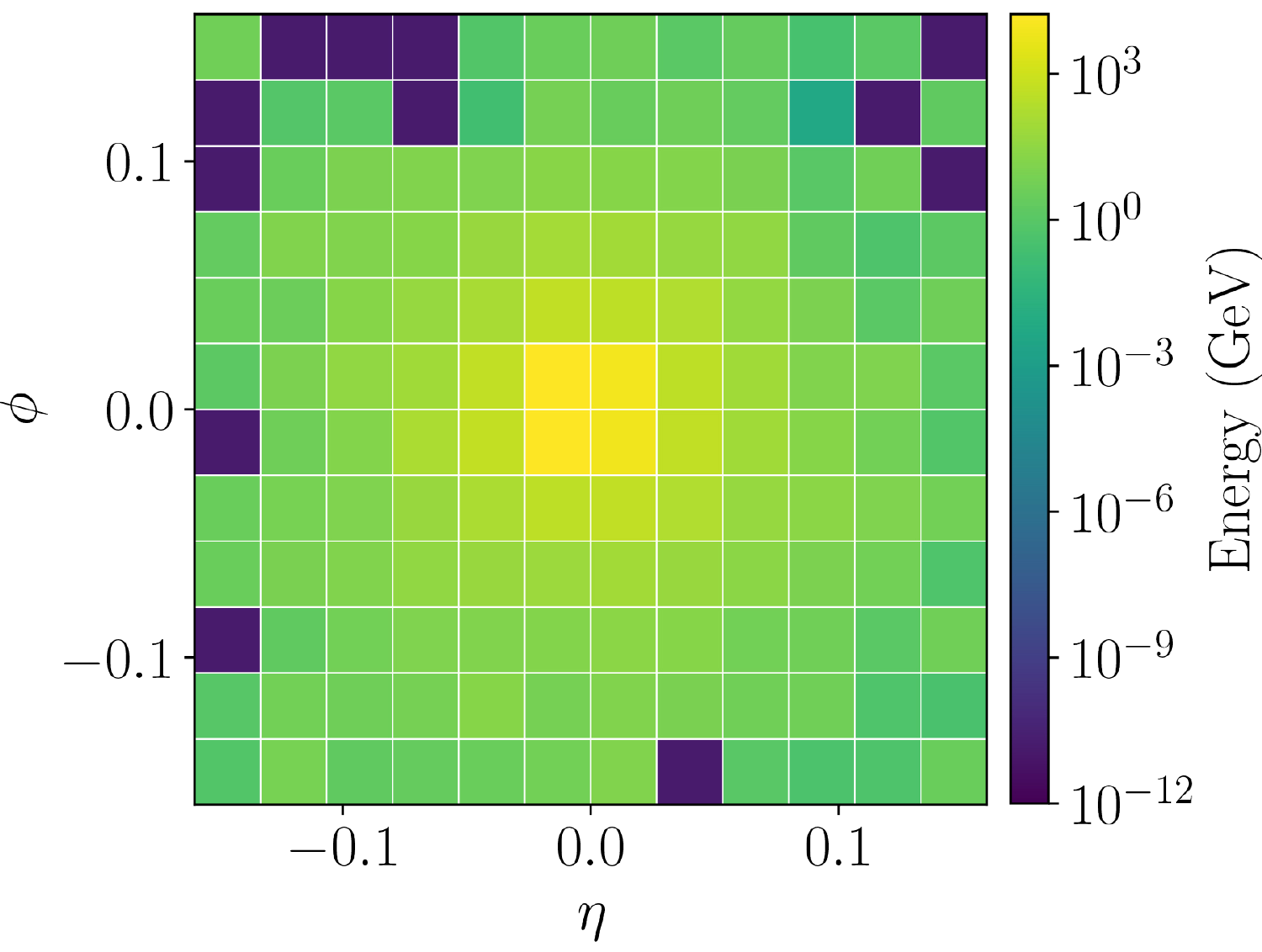}
        \caption{(12, 12)\\ }
        \label{fig:48x48}
    \end{subfigure}
    \hfill
    \begin{subfigure}[b]{0.24\textwidth}
        \centering
        \includegraphics[width=\textwidth]{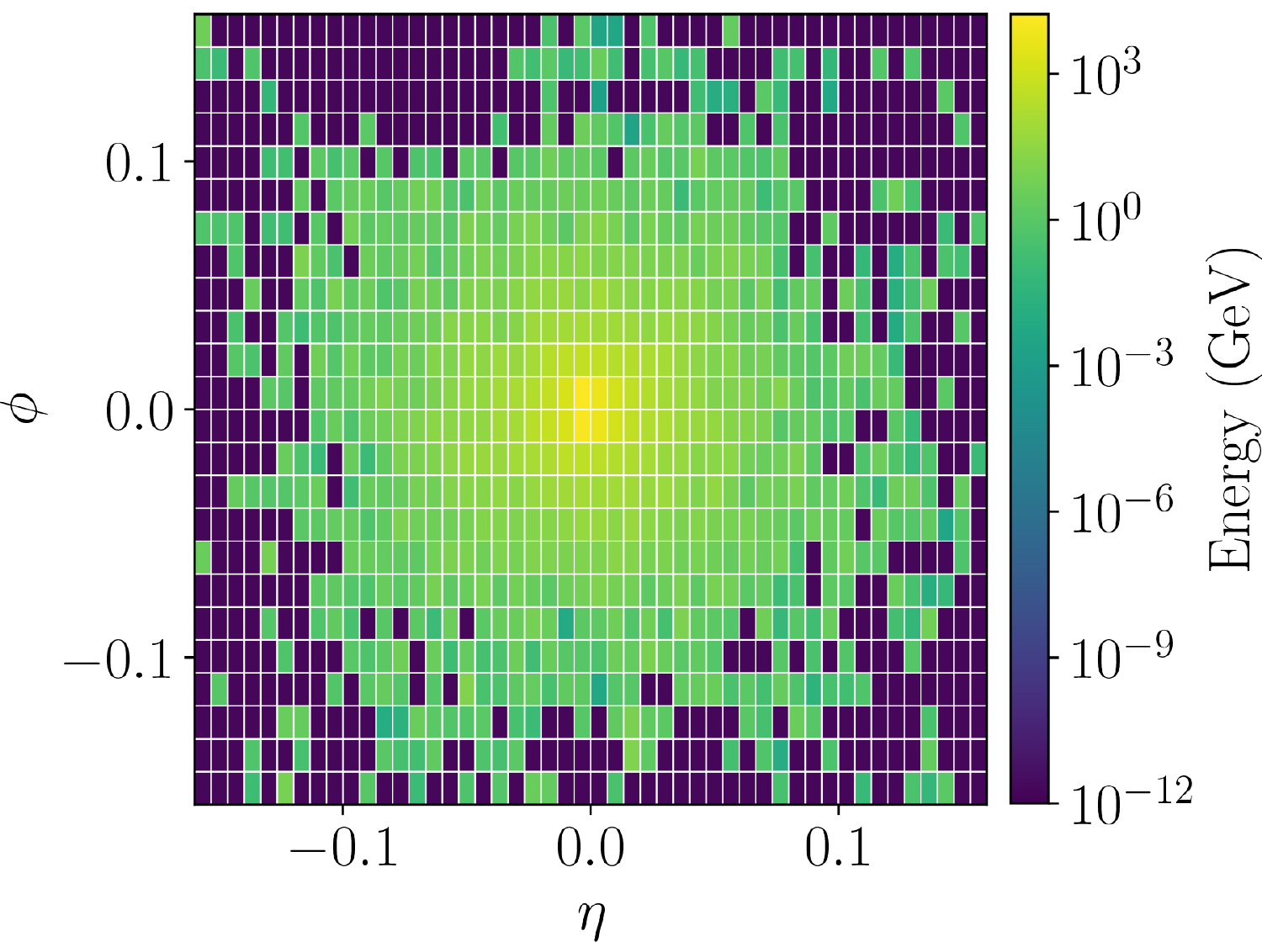}
        \caption{(48, 24)\\ }
        \label{fig:48x24}
    \end{subfigure}
    \hfill
    \begin{subfigure}[b]{0.24\textwidth}
        \centering
        \includegraphics[width=\textwidth]{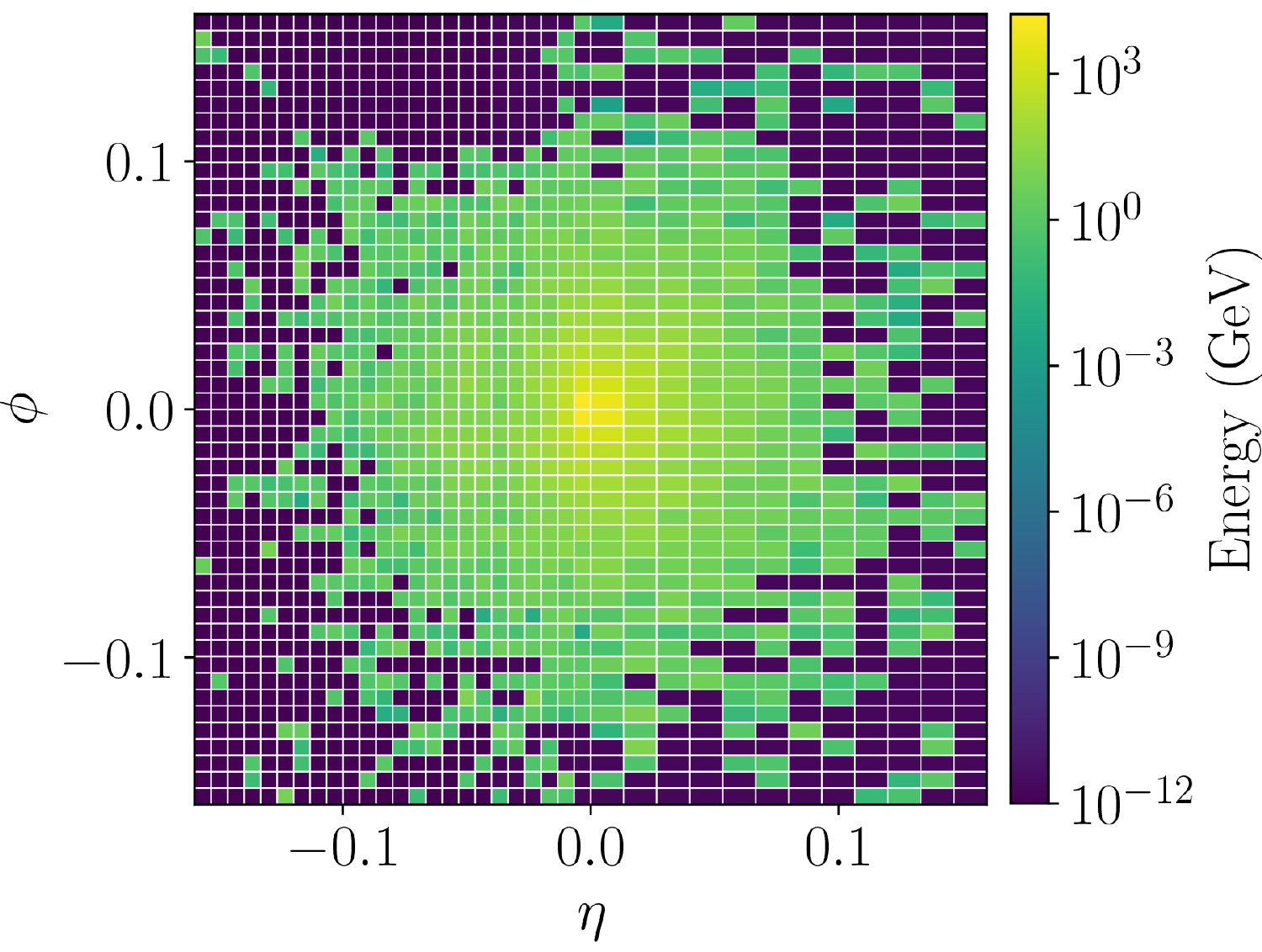}
        \caption{ Boundary Region}
        \label{fig:36x48}
    \end{subfigure}
    \caption{ Simulated calorimeter response to a 65 GeV photon for four distinct calorimeter cell segmentations.  Shown is the energy deposited per cell, for various segmentations in $(\eta,\phi)$. The Boundary Region is (24, 48)  for $\eta<0$ and (12, 48) for $\eta>0$.  }
    \label{fig:fourGeometries}
\end{figure}

\section{Context and Related Work}
\label{sec:relatedWork}

The ATLAS electromagnetic calorimeter has a complex geometry, and the simulation of its response is a computational bottleneck, motivating physicists to develop faster simulations. Early versions approximated the calorimeter response with parameterized response functions~\cite{ATLAS:2010bfa}. There has been significant innovation in developing DGMs for particle physics calorimeters using fixed output geometries, including with Generative Adversarial Networks (GAN)~\cite{Paganini:2017dwg, 3dgan_epj}, Variational Auto-Encoders (VAE)~\cite{Buhmann:2020pmy}, Normalizing Flows (NF)~\cite{Krause:2021ilc,Krause:2021wez}, and most recently diffusion models~\cite{Mikuni:2022xry}. 

Conditional generative models have been explored, such as an early effort by the ATLAS collaboration~\cite{ATLAS:2022jhk}, which attempts to condition a GAN on the entire calorimeter geometry at once. This ambitious effort faces the challenge of the enormous number of possible three-dimensional geometries, which requires generating a vast  training dataset in Geant4, and presents difficulties in conditioning the GAN.  
On the other hand, training hundreds of GANs~\cite{ATLAS:2021pzo} for small sections of the calorimeter with fixed geometries raises questions about the sustainability of validating and maintaining so many models. Experiments like CMS and LHCb have also studied DGMs for their calorimeters~\cite{Erdmann:2018jxd,Ratnikov:2020dcm}, and related studies have been made for non-LHC experiments~\cite{Hashemi:2023ruu}.

Instead of conditioning a DGM on all possible unique geometries, it is more natural to condition it on the properties of individual calorimeter cells. In this way, the model can be expected to interpolate between different cell sizes and positions that it has encountered in training,  and learn from multiple cells in the same image. An autoregressive algorithm, which generates cells sequentially, provides an efficient solution by conditioning the model solely on the properties of the cell being generated and its neighbors. This cell-by-cell approach enables the model to learn more efficiently how the properties of each cell affect its energy distribution. Recently, Sparse Autoregressive Models were studied for simulating sparse calorimeter images of a fixed geometry, in which the model learns the sparsity of the model with a tractable likelihood~\cite{lu2021sparse}. This model serves as the inspiration for our work. The details of the construction of the model are described in Sec.~\ref{sec:genModels}.

\section{Dataset}
\label{sec:Data}

 Samples of simulated electromagnetic calorimeter response are generated using Geant4~\cite{GEANT4:2002zbu,Allison:2016lfl,Allison:2006ve}, in an ATLAS-like configuration modified from that used to generate earlier datasets~\cite{Paganini_2018}.  The detector transverse segmentation (coordinates  $\eta$ and $\phi$) was modified, to significantly increase the granularity of the calorimeter, up to 16 fold, and approximate the granularity of potential future detectors. Several different  configurations were used to generate responses from a wide set of hypothetical geometries. These configurations were inspired by the most challenging geometry configurations seen in the ATLAS electromagnetic calorimeter~\cite{ATLAS:2008xda, Aleksa_Diemoz_2013}.
 
 The calorimeter comprises three longitudinal layers referred to as `inner,' `middle,' and `outer' layers and denoted by $z={1, 2, 3}$, each with area $48 \times 48$ cm$^2$. The layers have a length of 5mm, 40mm, and 80mm in $\eta$, 160mm, 40mm and 40mm in $\phi$ and a depth of 90mm, 347mm and 43mm respectively. For further details about the detector design, refer to Ref~\cite{Paganini_2018}. The calorimeter layers are segmented into various rectangular cells using a grid in $(n_{\eta},\ n_{\phi})$, the number of cells along the $\eta$ and $\phi$ directions, respectively; see Table~\ref{tab:geo_table}. In one particular configuration, the middle layer is structured with two different cell sizes to represent a boundary between detector regions which presents a difficult challenge for any generative model. 
In this non-uniform\footnote{All non-uniform geometries are marked with a *.} (36, 48)* geometry, the right half of the image is segmented into a (12, 48) grid and the left half into (24, 48) cells in $(n_{\eta},\ n_{\phi})$. The outermost layer had a consistent shape of (24, 24).
 
A sample of 10,000  simulated photons with energy of 65 GeV are directed at the  center of the calorimeter. The photon interaction with the material generates a shower of electrons and photons which deposit energy in each layer. The middle layer receives the largest fraction of the photon energy, while the outer receives the least. The simulated response to a photon for four example calorimeter geometries are shown in Fig.~\ref{fig:fourGeometries}, demonstrating the dramatic impact of the geometry. 

\begin{center}
\label{tab:geo_table}
\begin{tabular}{ll}
  \toprule
  \bf{Layer} & \bf{Calorimeter Segmentation} \\ 
  \midrule
  Inner & \makecell{(48,4), (48,12), (48,24), (48,48), \\
  (192,4), (192,12), (192,24), (192,48)}\\ 
  \midrule
  Middle & (12,12), (48, 24), (48,48), (36,48)* \\ 
  \midrule
  Outer & (24,24) \\ 
  \bottomrule
\end{tabular}
\captionof{table}{Segmentation of each layer of the simulated calorimeters used in training, indicated by $(n_{\eta},\ n_{\phi})$, the number of cells in $\eta$ and $\phi$, respectively. Several configurations are used, including one (marked with a *) in which the cell division is not uniform. See text for details.}
\end{center}

In addition, a second set of configurations is prepared and photon events simulated, but which is held out of the training set in order to test the generalization to new geometries.   Configurations with a middle layer segmentation of (24, 24) or (24, 12) are generated to test the ability of the model to interpolate. Configurations with a middle layer segmentation of (96,24), (6,6) or (9, 12)* are generated to test the ability of the model to extrapolate beyond its training range (summarised in Table~\ref{tab:interp_extrap_table}).

\begin{center}
\label{tab:interp_extrap_table}
\begin{tabular}{ll}
  \toprule
  \bf{Interpolation} & \bf{Extrapolation} \\ 
  \midrule
  (24,24), (24,12) & (96,24), (6,6), (9, 12)*\\ 
  \bottomrule
\end{tabular}
\captionof{table}{Segmentation of interpolation and extrapolation middle layer geometries of the simulated calorimeters used in evaluation, indicated by $(n_{\eta},\ n_{\phi})$, the number of cells in $\eta$ and $\phi$, respectively. Several configurations are used, including one (marked with a *) in which the cell division is not uniform. See text for details.}
\end{center}

\section{Generative Models}
\label{sec:genModels}

Generative models, both conditioned on the geometry and unconditioned, are developed.  Both are described below.

\subsection{Geometry-Aware Autoregressive Model (GAAM)}

\begin{figure}
  \centering
  \subfloat[GAAM]{\includegraphics[width=.74\textwidth]{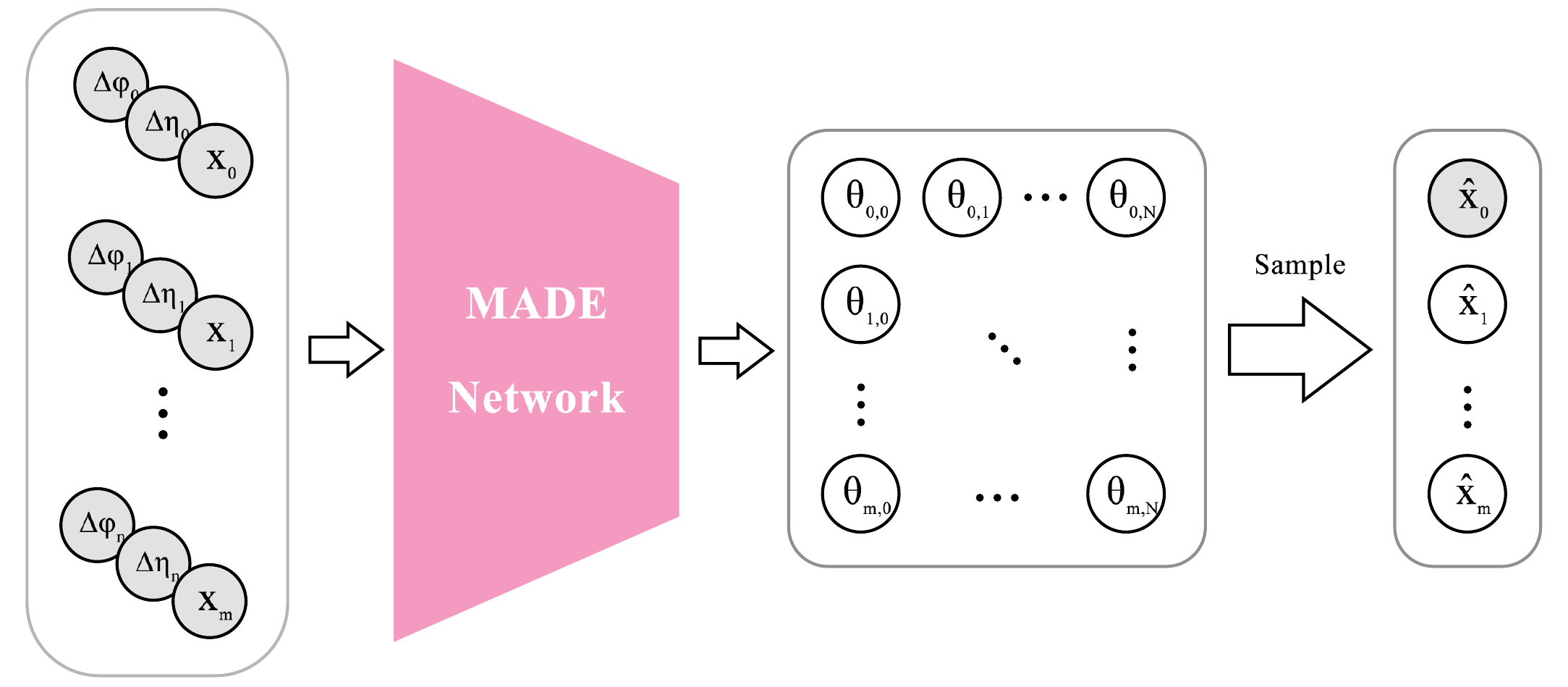}\label{withcellsize}}
  \subfloat[Spiral order]{\includegraphics[width=.26\textwidth]{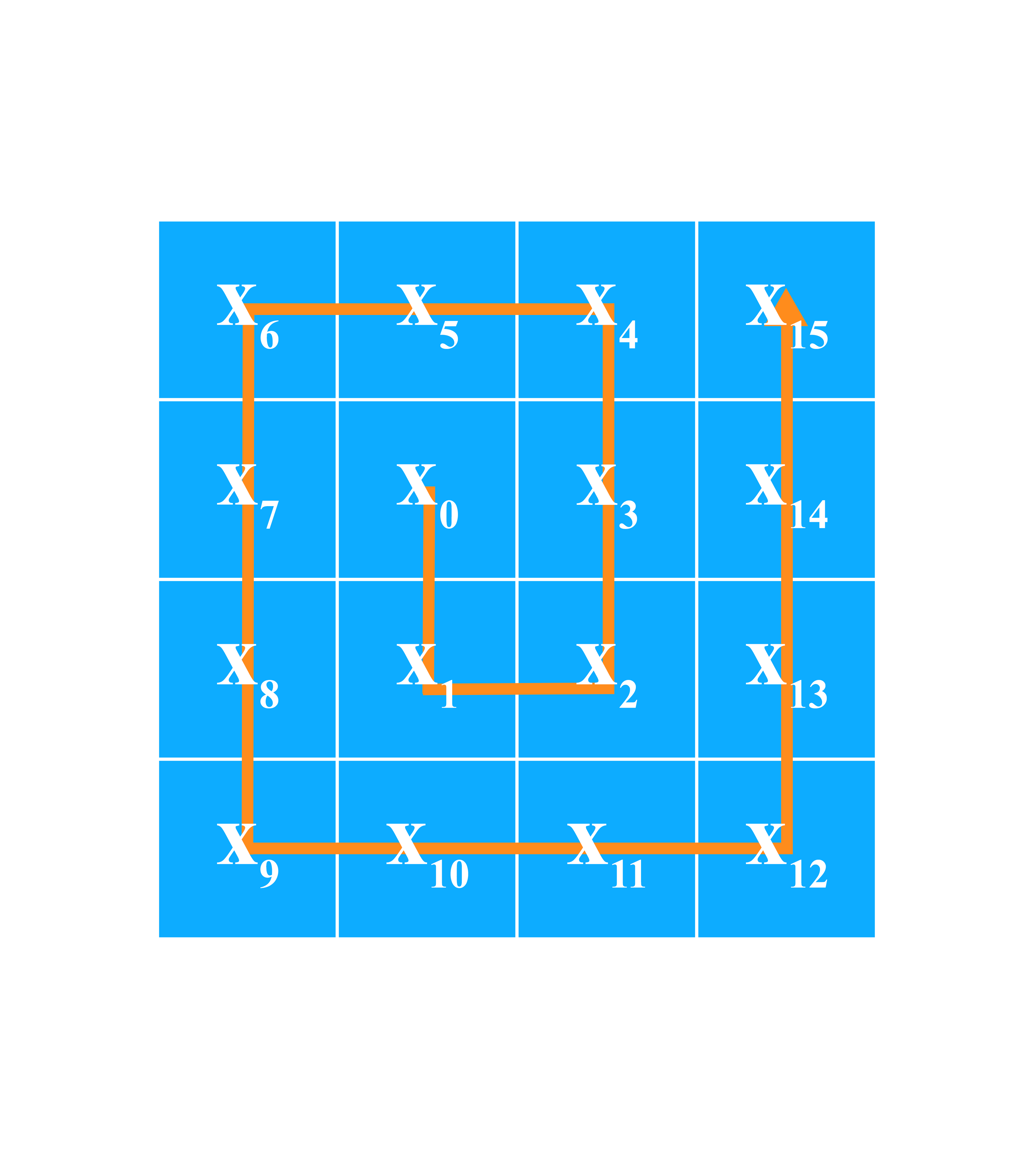}\label{spiralpath}}
  \caption{(a) 
The ARM architecture in GAAM takes a starting value $x_0$ and cell size ($\Delta\eta,\ \Delta\phi$) as inputs, which are aggregated to the MADE network. The parameters $\mathrm{\theta_{i}}$ represent the categorical distribution model (with $N+1$ discrete energy categories) that has been learned, and the $m - 1$ output calorimeter cell energies are sampled from this distribution.
  (b) The two-dimensional calorimeter layer matrix is flattened along a spiral path.}
  \label{architecture}
\end{figure}

The GAAM is a framework based on autoregressive models (ARM) that can efficiently learn to generate 3D calorimeter images. It comprises three ARMs, each generating an individual 2D layer of the calorimeter (see Figure~\ref{architecture}). Each ARM is constructed with Masked Autoencoder for Distribution Estimation (MADE)~\cite{germain2015made}. Unlike traditional autoregressive models that sequentially generate output cells depending on preceding cells, MADE generates all desired parameters in a single pass through the regular autoencoder. This enables parallel computation on GPUs for faster training.

To generate the inner calorimeter layer, an ARM is employed consisting of one masked fully-connected layer and one 1D convolutional layer. For the middle and outer calorimeter layers,  ARMs are used comprising five masked fully-connected layers and one 1D convolutional layer. Each of them uses GELU~\cite{hendrycks2016gaussian} activation functions. The output of the model is discrete energy (with $N+1$ possible values) of $m$ cells, and each generated cell energy deposit is represented as a categorical distribution (see Sec.~\ref{sec:Preprocessing} for the details about the preprocessing of cell energies). A softmax layer  is applied to generate these discrete outputs into $N+1$ categories, where $N$ is the closest integer greater than the maximum cell energy in the training data. 

In addition to the previous energy deposits, the ARM takes into account the cell sizes $(\Delta \eta,\ \Delta \phi)$ in the $\eta$ and $\phi$ dimensions, respectively, as conditioning features. This enables the model to dynamically generate the energy distribution to accommodate different geometries (Figure~\ref{withcellsize}). By learning to generate cell-by-cell, the model gains insight into how the spatial properties of each cell impact the energy deposition. For instance, the model should learn to deposit more energy per cell for a less granular geometry with bigger cells compared to a finer geometry with smaller cells, as well as take into account where the cell is in the image. Consequently, the model learns from multiple cells within each image, enhancing the learning efficiency. During the generation, each ARM receives a starting value, the  energy of a central cell. The energy of the central cell is sampled from a known prior distribution. The generation process is initiated from this cell since the majority of the energy tends to be deposited near the center, providing a reliable starting point for the ARMs. In addition to the central cell energy, the networks also receive two matrices describing the size of the cells to be generated. The trained GAAM then generates the energy deposits in the remaining cells by sampling from the learned categorical distribution.

\subsection{Baseline: Unconditional ARM}

As a comparison to the geometry-aware approach described above, a similar model is trained  without conditioning it to the properties of the calorimeter cells. This will serve as a baseline to understand the improvement that comes purely from being able to condition the model on cell properties.

\subsection{Preprocessing}
\label{sec:Preprocessing}
Since an ARM requires an ordering of the calorimeter cells, the two-dimensional cell matrix is flattened using a spiral counter-clockwise pattern (Figure~\ref{spiralpath}) so that  cells remain close after flattening. The ordering starts from a central cell, where most energy is deposited, and moves outwards. Given the range in the distribution of energy depositions across cells spans several orders of magnitude, applying a power transformation, $\hat{x}=x^{1/p}$, with $p=2$ during preprocessing is found to improve the training. Furthermore, the value is then discretized by rounding it to the nearest integer. 

The dataset is divided into training, validation, and test sets, with a ratio of 0.7:0.1:0.2, respectively. Training the models is carried out in PyTorch~\cite{NEURIPS2019_9015} for up to 200 epochs, using the Adam optimizer~\cite{kingma2014adam}. The negative log likelihood serves as the loss function. The training is performed on two NVIDIA RTX 3090 GPUs.

\section{Results}
\label{sec:results}

Performance of the conditional GAAM as well as the baseline unconditional model are presented both qualitatively as well as quantitatively, for geometries with uniform or non-uniform cell segmentation, and for geometries within the training sample as well as unseen geometries.

Qualitative analysis is via visualizations and inspection of generated samples. Quantitative analysis is performed by measuring the distance between histograms of true and generated samples in quantities which are important in a typical data analysis: energy weighted means, shower widths, and distances between shower means across layers. The energy weighted means, $\bar{\eta}$ and $\bar{\phi}$ are computed over all cells in an image:
\begin{equation}\footnotesize
\label{eq_energy_weighted_mean}
\bar{\eta} = \frac{\sum_{i} \eta_{i}E_{i}}{\sum_{i} E_{i}},  \bar{\phi} = \frac{\sum_{i} \phi_{i}E_{i}}{\sum_{i} E_{i}} 
\end{equation}
where $E_{i}$ is the energy deposited in the $i^{th}$ cell. The shower widths, $\sigma_{\eta}$ and $\sigma_{\phi}$ are calculated as:
\begin{equation}\footnotesize
\label{eq_shower_width}
\sigma_{\eta} = \sqrt{\frac{\sum_i E_{i} (\eta_i-\bar{\eta})^2}{\frac{(M-1)}{M} \sum_{i} E_{i}}}, \sigma_{\phi} = \sqrt{\frac{\sum_i E_{i} (\phi_i-\bar{\phi})^2}{\frac{(M-1)}{M} \sum_{i} E_{i}}}
\end{equation}
where $M$ is the number of cells with non-zero energy. The inter-layer distance between mean energy-weighted shower locations in the inner and middle calorimeter layers are computed as: 
\begin{equation}\footnotesize
    C_{\eta} = \bar{\eta}_\textrm{inner}-\bar{\eta}_\textrm{middle}, \ \ 
    C_{\phi} = \bar{\phi}_\textrm{inner}-\bar{\phi}_\textrm{middle} 
\end{equation}
Distances between histograms are measured using the Wasserstein distance~\cite{wasserstein_original,villani_ot}. The speed of the GAAM is a function of the number of cells. Generation takes 2040.7 ms/shower for inner layers, 140.5 ms/shower for the middle and 9.6 for outer layers, compared to  2058 ms/shower with Geant4 for 65 GeV photons. However, this model had not yet been optimized for speed.

\subsection{Performance in uniform segmentation}

\begin{figure}[t!]
  \centering
  \subfloat[GAAM - Inner layer - (192, 12)]{\includegraphics[width=.45\textwidth]{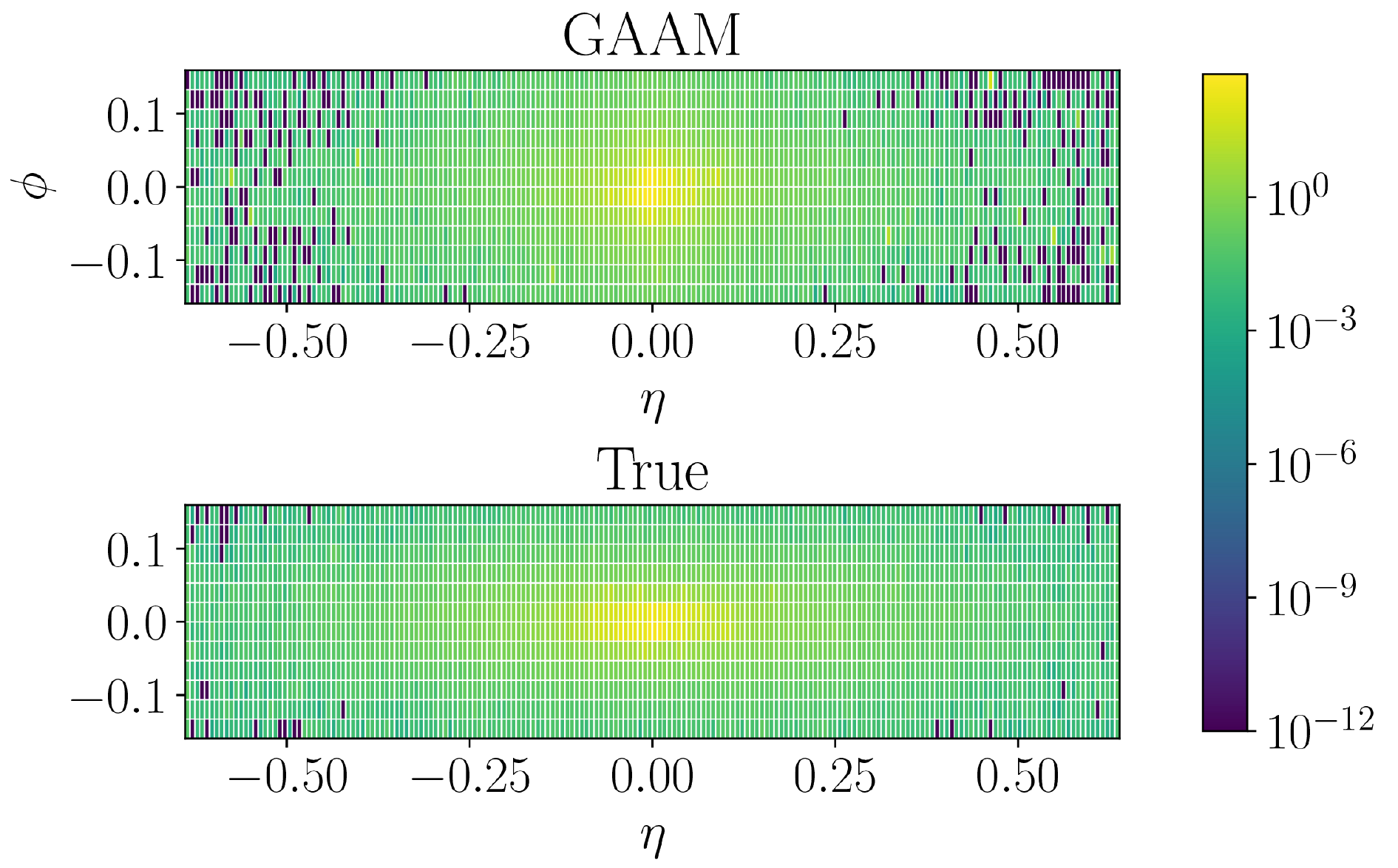}\label{meanimage_stripelayer_192x12}}\hfill
  \subfloat[GAAM - Middle layer - (36, 48)*]{\includegraphics[width=.45\textwidth]{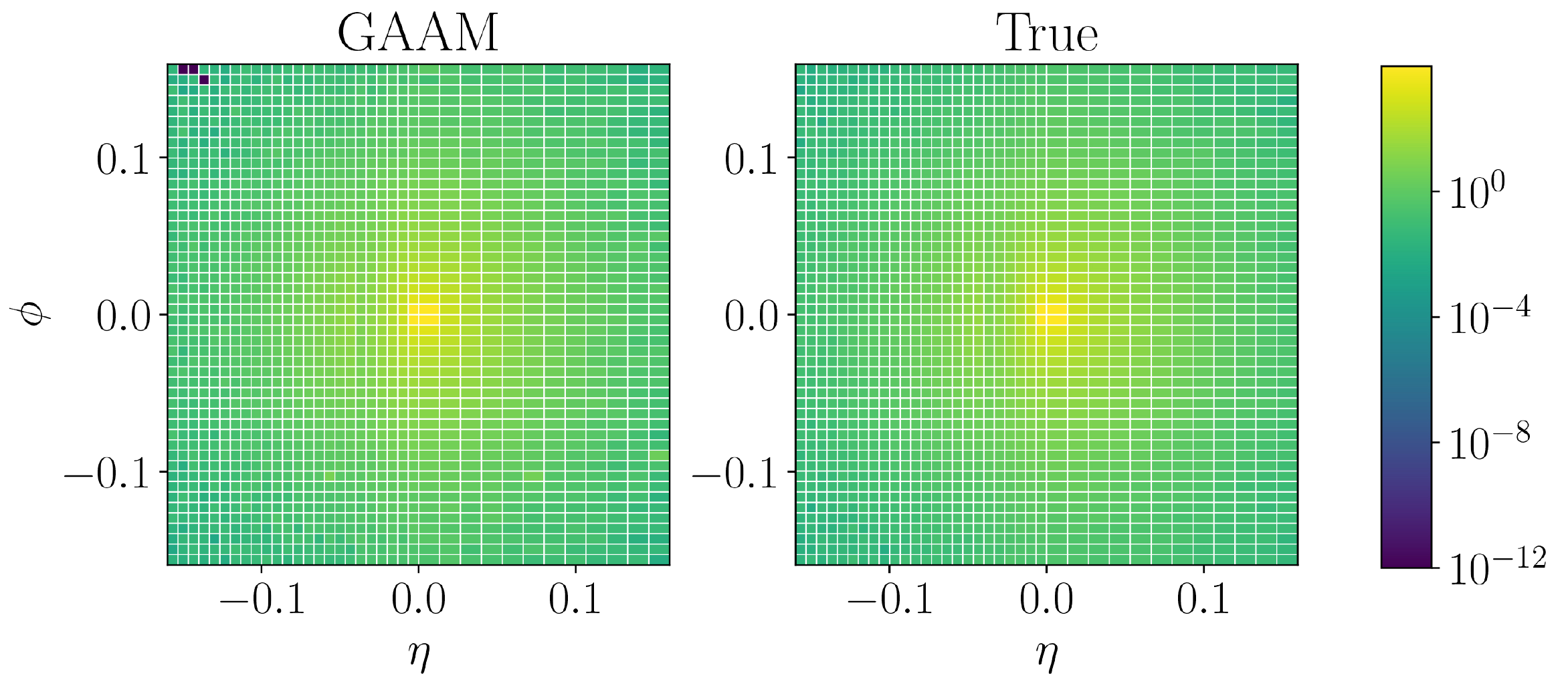}\label{meanimage_middlelayer_36x48}}\hfill
  \subfloat[GAAM - Outer layer - (24, 24)]{\includegraphics[width=.45\textwidth]{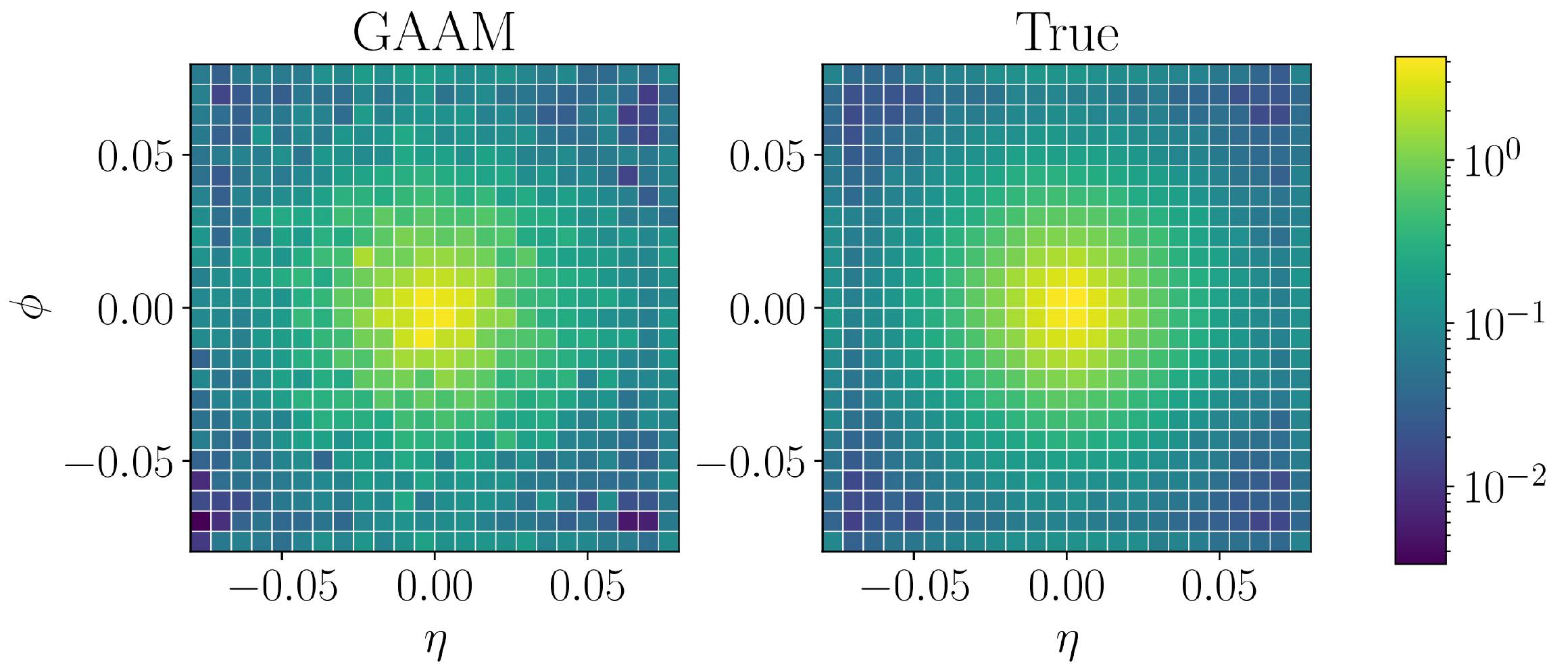}\label{meanimage_middlelayer_24x24}}\hfill
  \subfloat[Unaware - Middle layer - (36, 48)*]
  {\includegraphics[width=.45\textwidth]{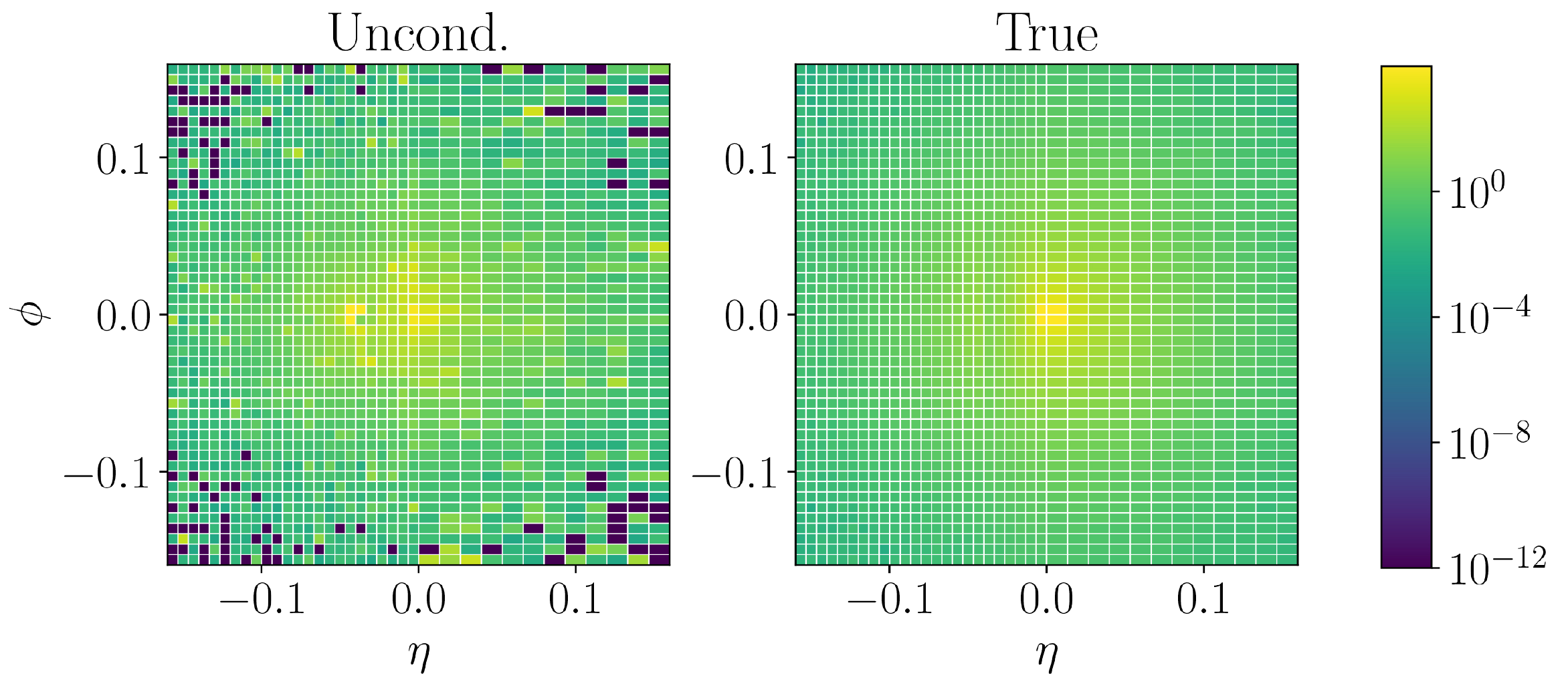}\label{meanimage_middlelayer_baseline_36x48}}\hfill
  \caption{Average generated calorimeter images from GAAM in (a) inner  (b) middle or (c) outer layers, or (d) from the geometry-unaware model middle layer, each compared the true images from Geant4.}
  \label{average_image}
\end{figure}

Qualitative comparison of the generated images from the GAAM and the geometry-unaware model are shown in Fig.~\ref{average_image} in each layer, for several example segmentations, and compared to the truth generated by Geant4. The GAAM reproduces the features of the true images, while the geometry-unaware model fails to do so.

\begin{figure}[t!]
  \centering
  \subfloat[Inner layer - (192, 12)]{\includegraphics[width=.5\textwidth]{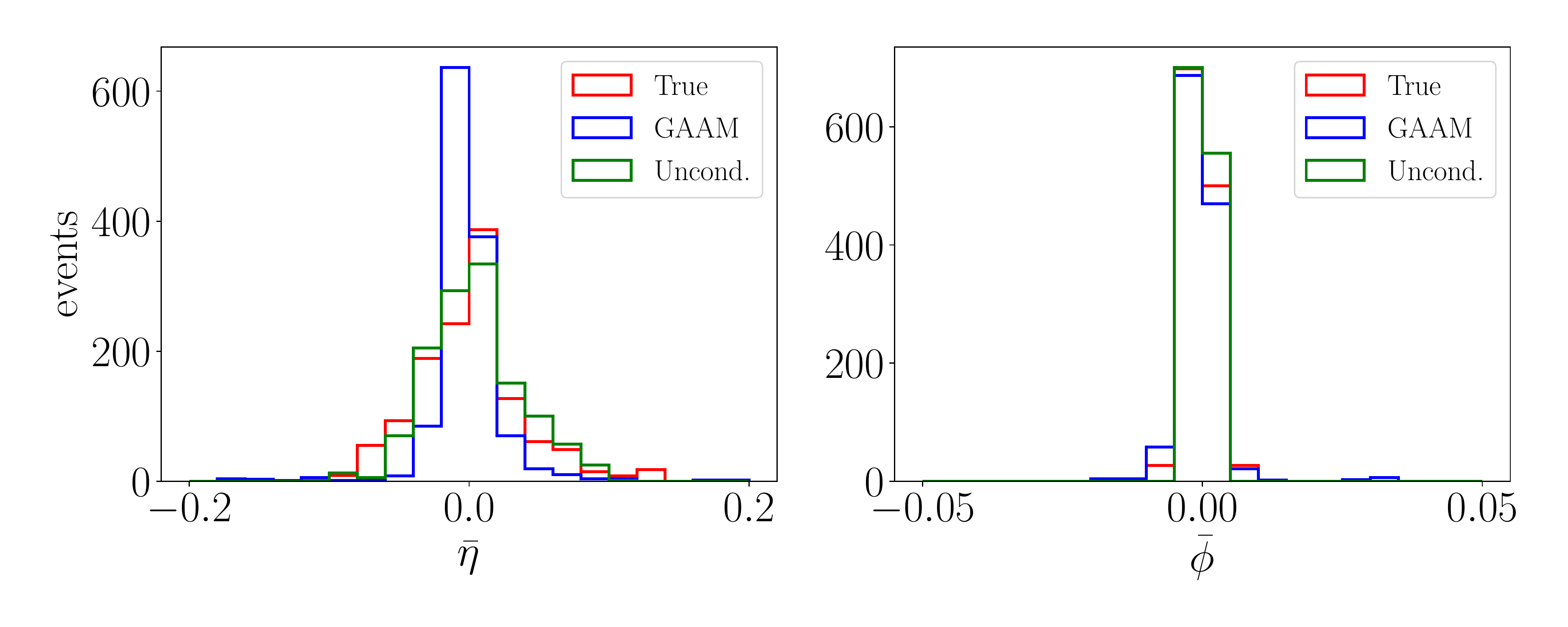}\label{eng-weighted_avgs_innerlayer_192x12}}\hfill
  \subfloat[Middle layer - (36, 48)*]{\includegraphics[width=.5\textwidth]{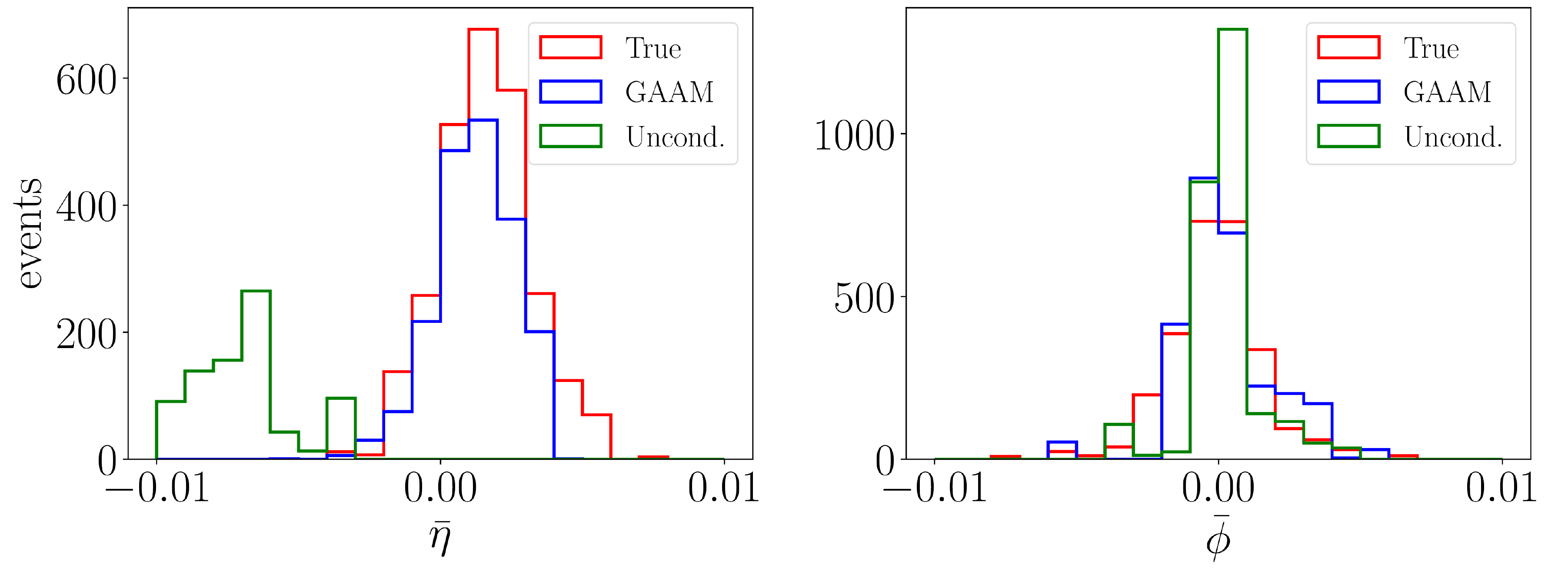}\label{eng-weighted_avgs_middlelayer_36x48}}\hfill
  \subfloat[Middle layer - (48, 48)]{\includegraphics[width=.5\textwidth]{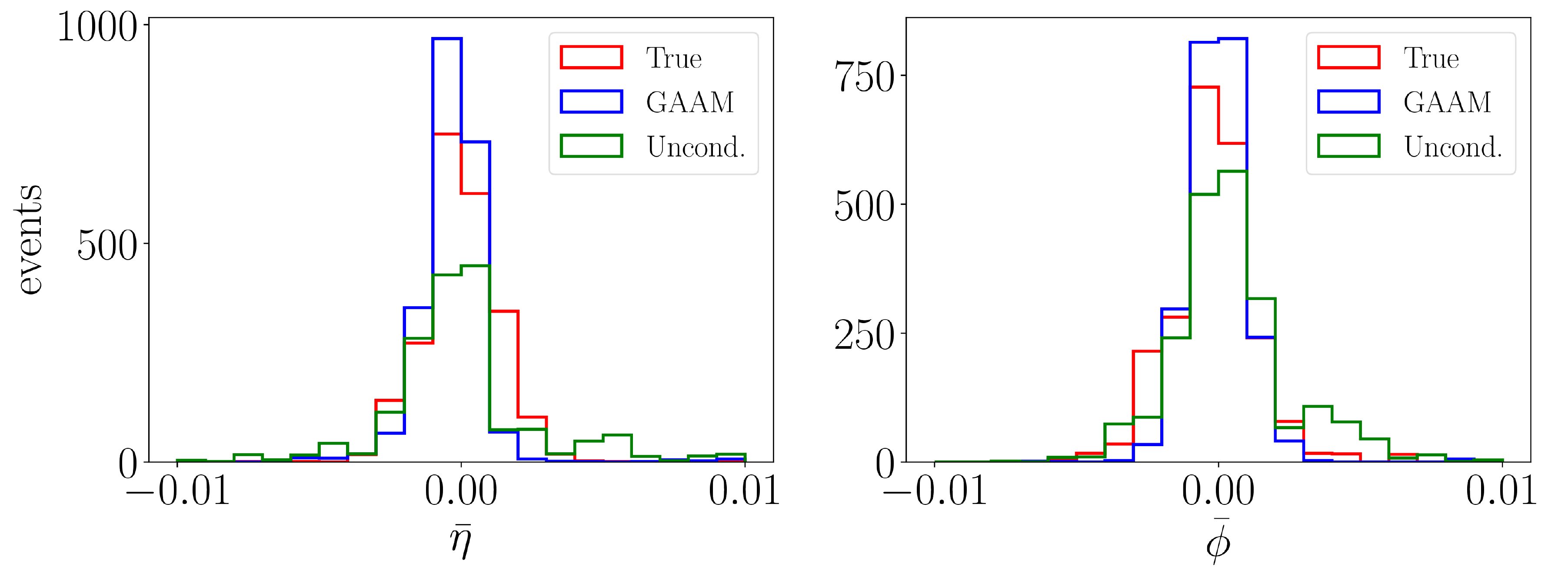}\label{eng-weighted_avgs_middlelayer_48x48}}\hfill
  \subfloat[Outer layer - (24, 24)]{\includegraphics[width=.49\textwidth]{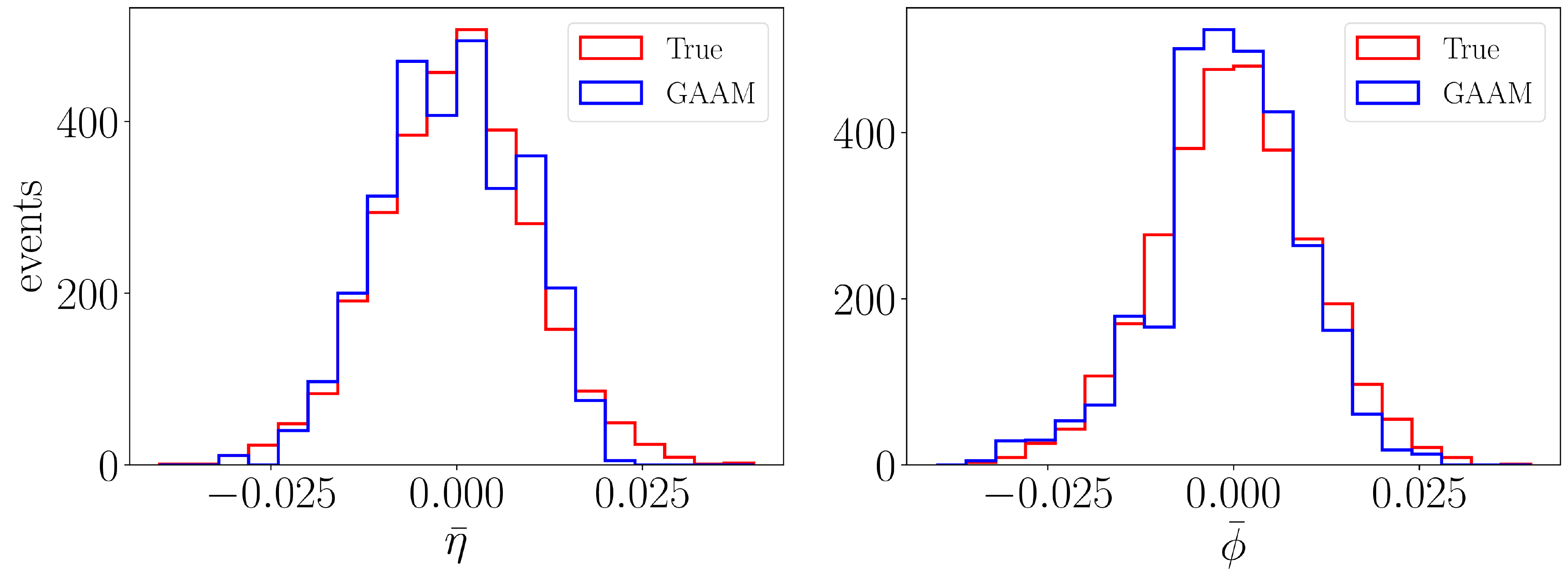}\label{eng-weighted_avgs_outerlayer_24x24}}\hfill
  \caption{Distribution of the energy weighted means, $\bar \eta$ and $\bar \phi$, in samples generated by the GAAM, the unconditional model and  Geant4 (True), for several calorimeter layers with varying segmentation in $(\eta,\ \phi)$.  The GAAM outperforms the unconditional  baseline model  consistently.}
  \label{energy-weighted_avgs}
\end{figure}

\begin{figure}[h!]
  \centering
  \subfloat[Inner layer - (192, 12)]{\includegraphics[width=.5\textwidth]{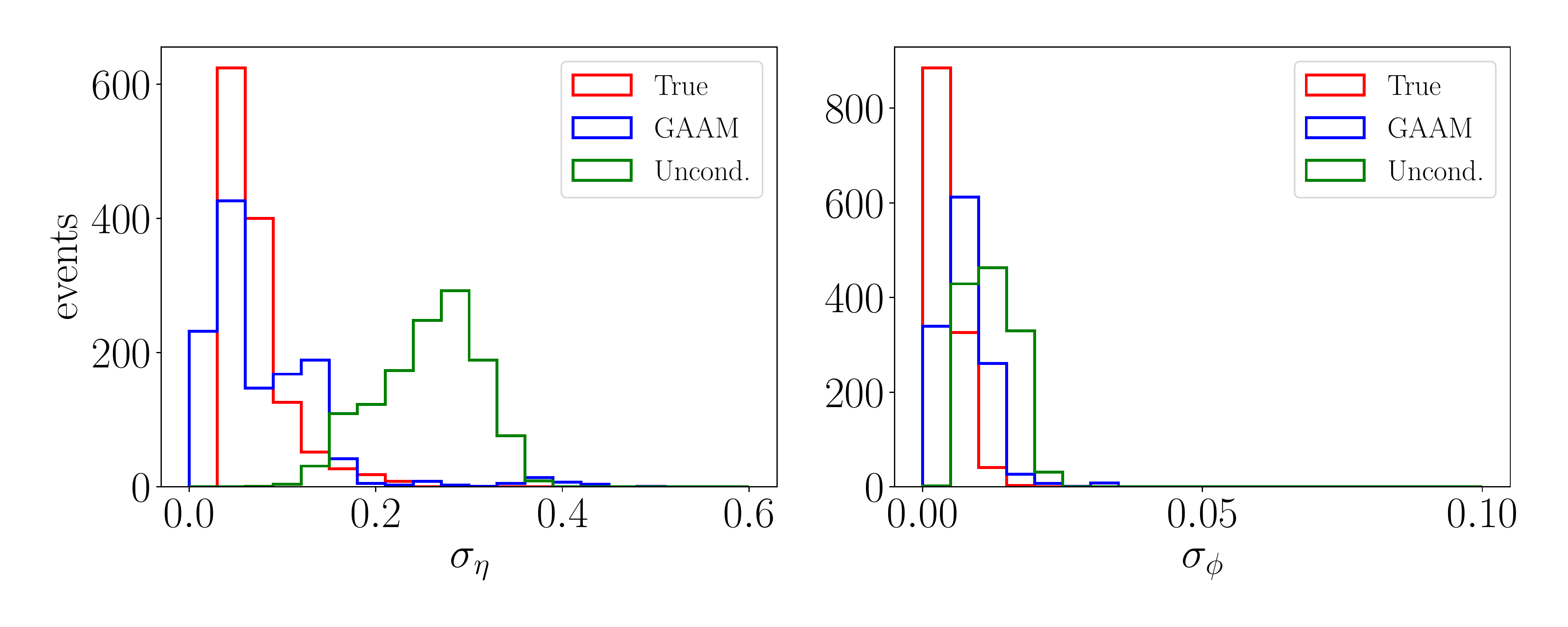}\label{shower_widths_stripelayer_192x12}}\hfill
  \subfloat[Middle layer - (36, 48)*]{\includegraphics[width=.5\textwidth]{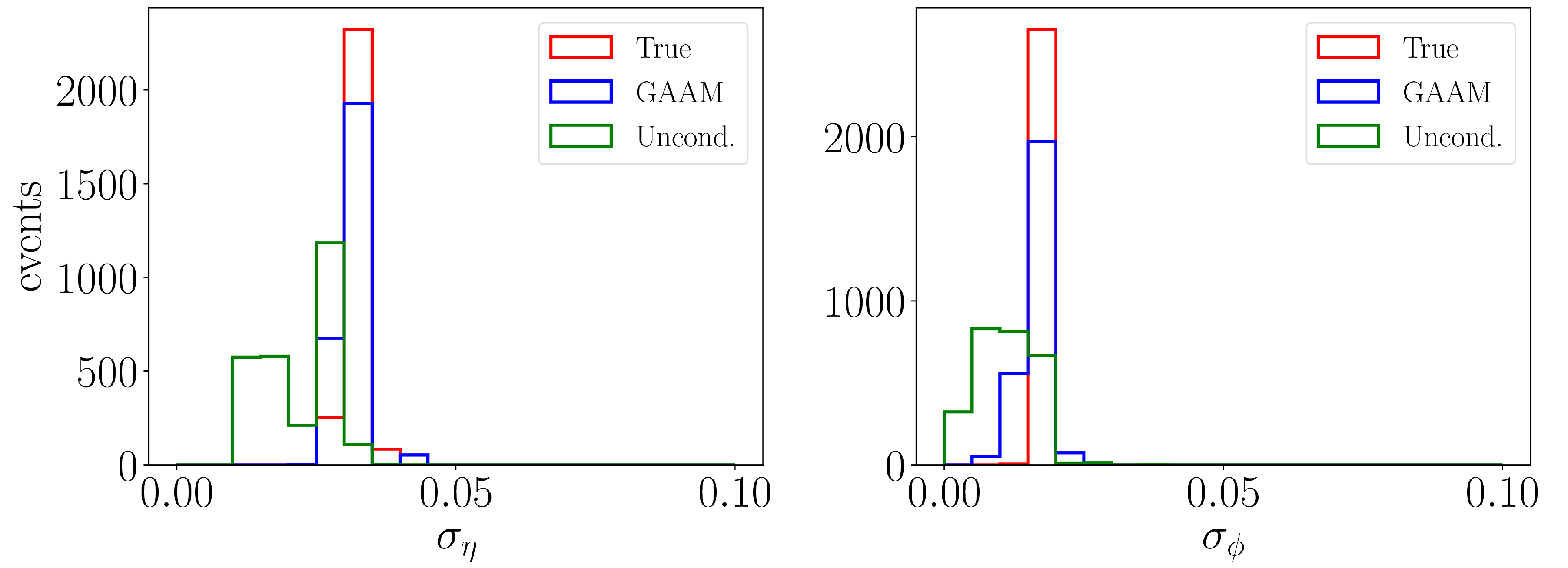}\label{shower_widths_middlelayer_36x48}}\hfill\\
  \subfloat[Middle layer - (48, 48)]{\includegraphics[width=.5\textwidth]{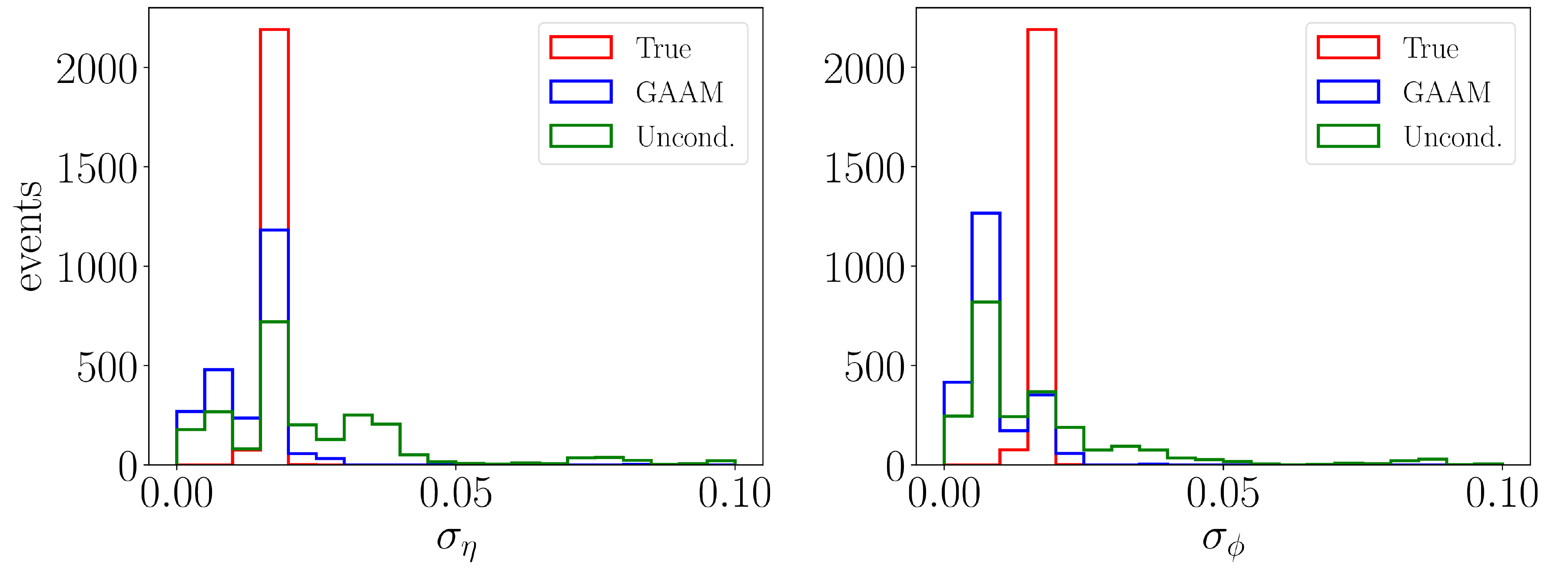}\label{shower_widths_middlelayer_48x48}}\hfill
  \subfloat[Outer layer - (24, 24)]{\includegraphics[width=.5\textwidth]{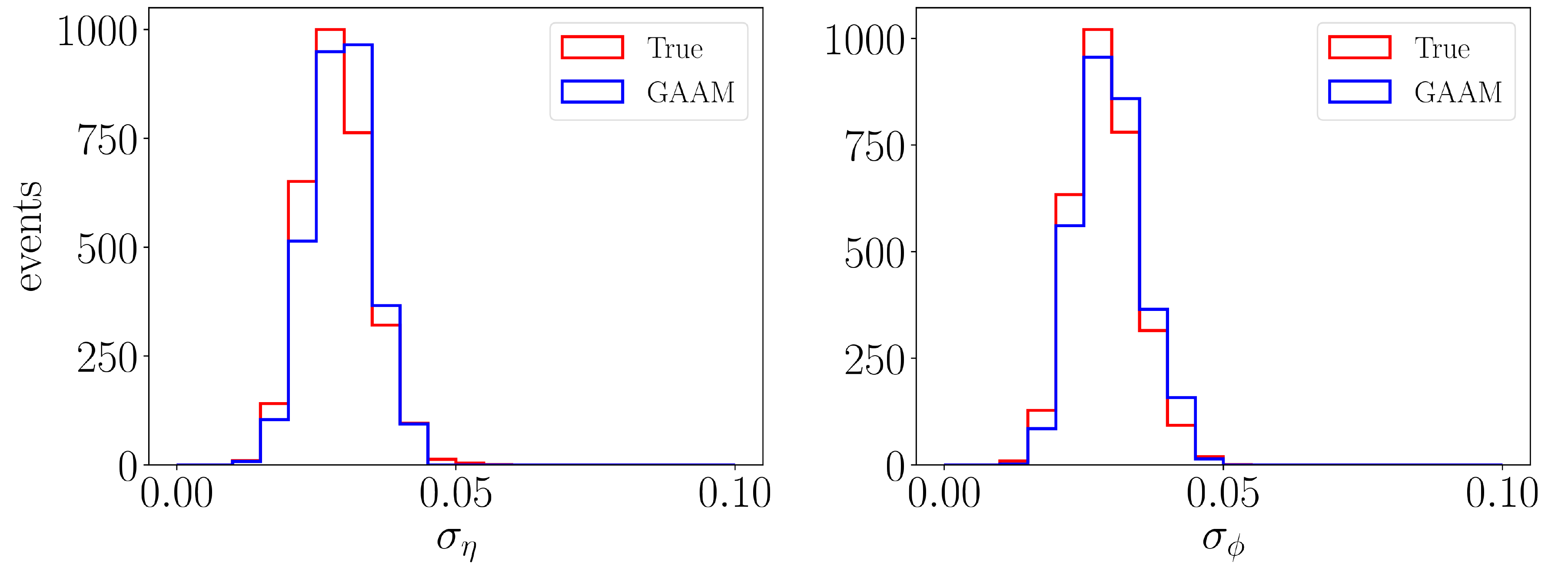}\label{shower_widths_middlelayer_24x24}}\hfill
  \caption{Distribution of the shower widths, $\sigma_{\eta}$ and $\sigma_{\phi}$, in samples generated by the GAAM, the unconditional model and  Geant4 (True), for several calorimeter layers with varying segmentation in $(\eta,\ \phi)$. The GAAM outperforms the unconditional  baseline model  consistently.}
  \label{shower_widths}
\end{figure}

\begin{figure}[t!]
  \centering
  \subfloat[Inner (48, 12) \& middle (12, 12) layer]{\includegraphics[width=.5\textwidth]{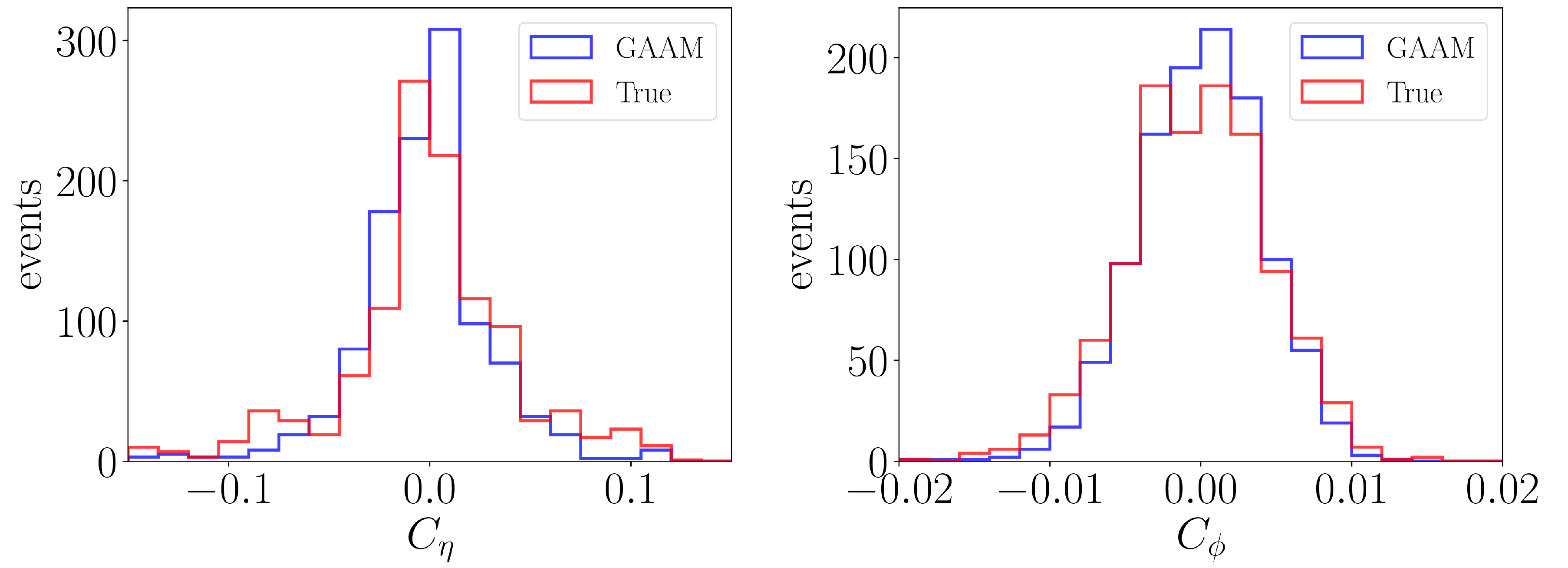}\label{shower_widths_middlelayer_48x12}}\hfill
  \subfloat[Inner (192, 48) \& middle (36, 48)* layer]{\includegraphics[width=.5\textwidth]{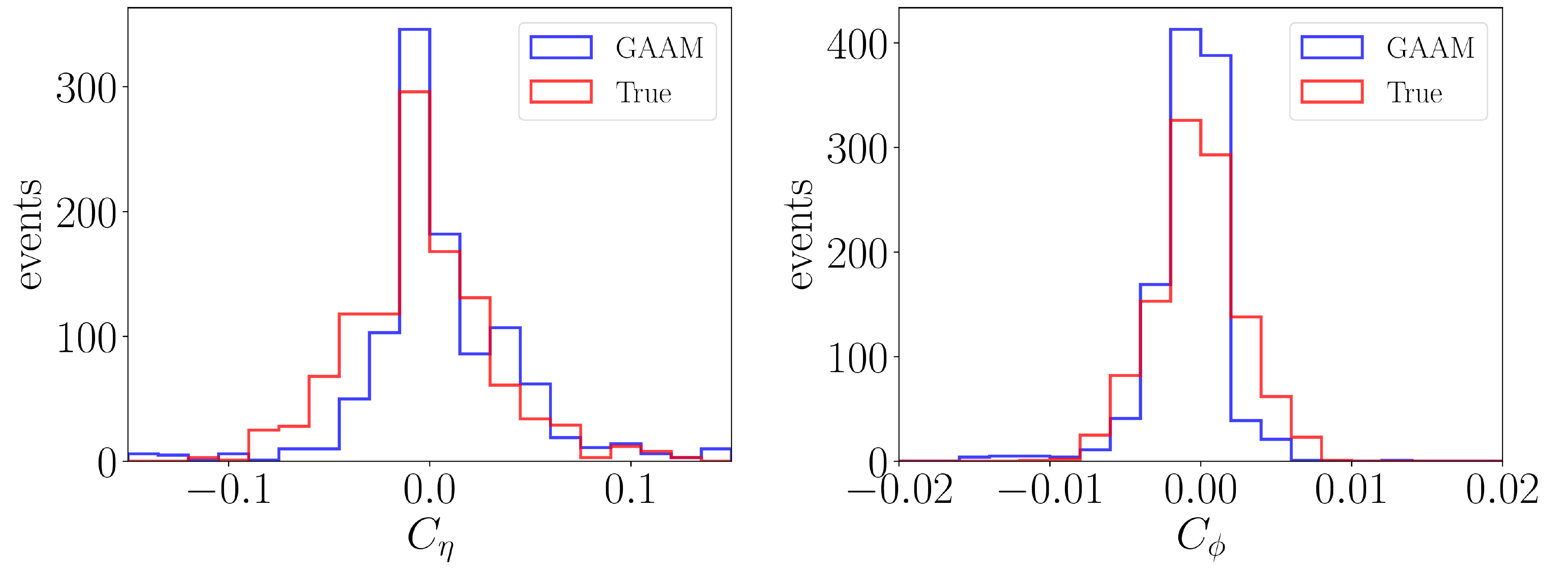}\label{shower_widths_middlelayer_192x48}}\hfill
  \caption{Distribution of the inter-layer distance of energy weighted means, $C_{\eta}$ and $C_{\phi}$,  in samples generated by the GAAM and  Geant4 (True), for several calorimeter layers with varying segmentation in $(\eta,\ \phi)$. The GAAM reproduces the truth reasonably well.}
  \label{layer_correlation}
\end{figure}

Quantitative comparison are made between histograms of energy-weighted means (Fig.~\ref{energy-weighted_avgs}), shower widths (Fig.~\ref{shower_widths}), and inter-layer distances (Fig.~\ref{layer_correlation}).  The Wasserstein distances between the distributions are given in Table~\ref{wdist_train}. In each case, the GAAM outperforms the geometry-unaware model\footnote{The outer layer has only a single geometry so no comparisons are made to a geometry-unaware model.}, and achieves a small inter-layer distance. The inner layer is particularly challenging for GAAM due to its high granularity.

\subsection{Performance in non-uniform segmentation}



A major challenge for implementing generative models is the non-uniform nature of the detector segmentation, which becomes more coarse or irregular in some regions. Performance under non-uniform cell divisions is studied in the configuration where the middle layer has segmentation (36, 48)*, but non-uniform cell sizes  along  $\eta$.  
Demonstration of the GAAM's ability to learn to generate data with larger energy deposition in larger cells is shown in Fig~\ref{meanimage_middlelayer_36x48}. The slight off-set of the peak to the right in the $\eta$ distribution in Fig.~\ref{eng-weighted_avgs_middlelayer_36x48} is a feature of boundary regions of calorimeters and therefore expected for the geometry (36, 48)* of the middle layer.

Such a task has never been attempted before with generative models, due to the challenges it poses, as illustrated by the failure of the geometry-unaware model to accurately simulate the calorimeter response in this scenario; see Figure~\ref{meanimage_middlelayer_baseline_36x48}.

\begin{table}
  \centering
  \begin{tabular}{llll}
    \toprule
    Layer; segmentation   & Quantity   & \multicolumn{2}{c}{Wasserstein distance}\\
    & &     GAAM (Conditional) & Baseline (Unconditional)\\
    \midrule
    \multirow{4}{*}{Inner; (192, 12)} & $\bar \eta$  & $1.80 \times 10^{-2}$  & $1.30 \times 10^{-2}$ \\  
                                        & $\bar \phi$ & $8.50 \times 10^{-4}$ & $1.36 \times 10^{-3}$ \\
                                        & $\sigma_{\eta}$ & $2.44 \times 10^{-2}$ & $1.63 \times 10^{-1}$ \\
                                        & $\sigma_{\phi}$ & $2.71 \times 10^{-3}$ & $6.45 \times 10^{-3}$ \\
    \midrule 
    \multirow{4}{*}{Middle; (48,48)} & $\bar \eta$  & $5.95 \times 10^{-4}$  & $6.98 \times 10^{-3}$ \\  
                                        & $\bar \phi$ & $4.05 \times 10^{-4}$  & $2.61 \times 10^{-3}$ \\
                                        & $\sigma_{\eta}$ & $4.35 \times 10^{-3}$ & $1.25 \times 10^{-2}$\\
                                        & $\sigma_{\phi}$ & $7.07 \times 10^{-3}$ & $1.00 \times 10^{-2}$ \\
    \midrule                                    
    \multirow{4}{*}{Middle; (36,48)} & $\bar \eta$  & $1.25 \times 10^{-2}$  & $2.45 \times 10^{-2}$\\  
                                        & $\bar \phi$ & $4.54 \times 10^{-4}$  & $5.58 \times 10^{-4}$ \\
                                        & $\sigma_{\eta}$ & $1.18 \times 10^{-3}$ & $1.00 \times 10^{-2}$ \\
                                        & $\sigma_{\phi}$ & $1.24 \times 10^{-3}$ & $5.60 \times 10^{-3}$ \\
    \midrule
    \multirow{4}{*}{Outer; (24,24)} & $\bar \eta$  & $8.31 \times 10^{-4}$  & N/A \\
                                        & $\bar \phi$ & $1.12 \times 10^{-3}$  & N/A \\
                                        & $\sigma_{\eta}$ & $9.08 \times 10^{-4}$ & N/A \\
                                        & $\sigma_{\phi}$ & $9.71 \times 10^{-4}$ & N/A \\
    \bottomrule
  \end{tabular}
  \caption{Quantitative evaluation of performance of the two generative models, the conditional GAAM and the unconditional baseline model in generating simulated calorimeter response to a photon. Shown is the Wasserstein distance between histograms of quantities important in data analysis, displayed in Figs.~\ref{energy-weighted_avgs}-\ref{layer_correlation}. The testing geometry appears in the training set for these models.}
  \label{wdist_train}
\end{table}

\begin{table}
  \centering
  \begin{tabular}{llll}
    \toprule
 Layer; segmentation   & Quantity   & \multicolumn{2}{c}{Wasserstein distance}\\
    & &     GAAM (Conditional) & Baseline (Unconditional)\\  
    \midrule
    \multirow{4}{*}{Middle; (24,24)} & $\bar \eta$  & $2.90 \times 10^{-4}$   & $7.74 \times 10^{-3}$ \\  
                                        & $\bar \phi$ & $6.21 \times 10^{-4}$    & $1.06 \times 10^{-3}$ \\
                                        & $\sigma_{\eta}$ & $8.98 \times 10^{-3}$ & $1.63 \times 10^{-2}$ \\
                                        & $\sigma_{\phi}$ & $4.22 \times 10^{-3}$ & $7.39 \times 10^{-3}$ \\
    \midrule
    \multirow{4}{*}{Middle; (24,12)} & $\bar \eta$  & $4.77 \times 10^{-4}$  & $7.74 \times 10^{-3}$ \\  
                                        & $\bar \phi$ & $1.02 \times 10^{-3}$ & $1.06 \times 10^{-3}$ \\
                                        & $\sigma_{\eta}$ & $8.98 \times 10^{-3}$ & $1.63 \times 10^{-2}$ \\
                                        & $\sigma_{\phi}$ & $4.22 \times 10^{-3}$ & $7.39 \times 10^{-3}$ \\

    \bottomrule
  \end{tabular}
  \caption{Quantitative evaluation of performance of the two generative models, the conditional GAAM and the unconditional baseline model in generating simulated calorimeter response to a photon. Shown is the Wasserstein distance between histograms of quantities important in data analysis, displayed in Fig.~\ref{new_shape_energy-weighted_avgs}. The testing geometry does not appear in the training set for these models, but requires interpolation.}
  \label{wdist_interp}
\end{table}

\subsection{Performance in unseen geometries}

A crucial target of geometry-aware models is the ability to learn the dependence of the energy deposition on the nature of the geometry and thus be able to  generate simulated response for calorimeters whose geometry do not appear in its training set. To explore this ability of the GAAM, we evaluate its performance on unseen geometries which require interpolation within the training set, as well as those which require extrapolation out of the training set.

\begin{figure}[t!]
  \centering
  \subfloat[Middle layer - (24, 24)]{\includegraphics[width=.5\textwidth]{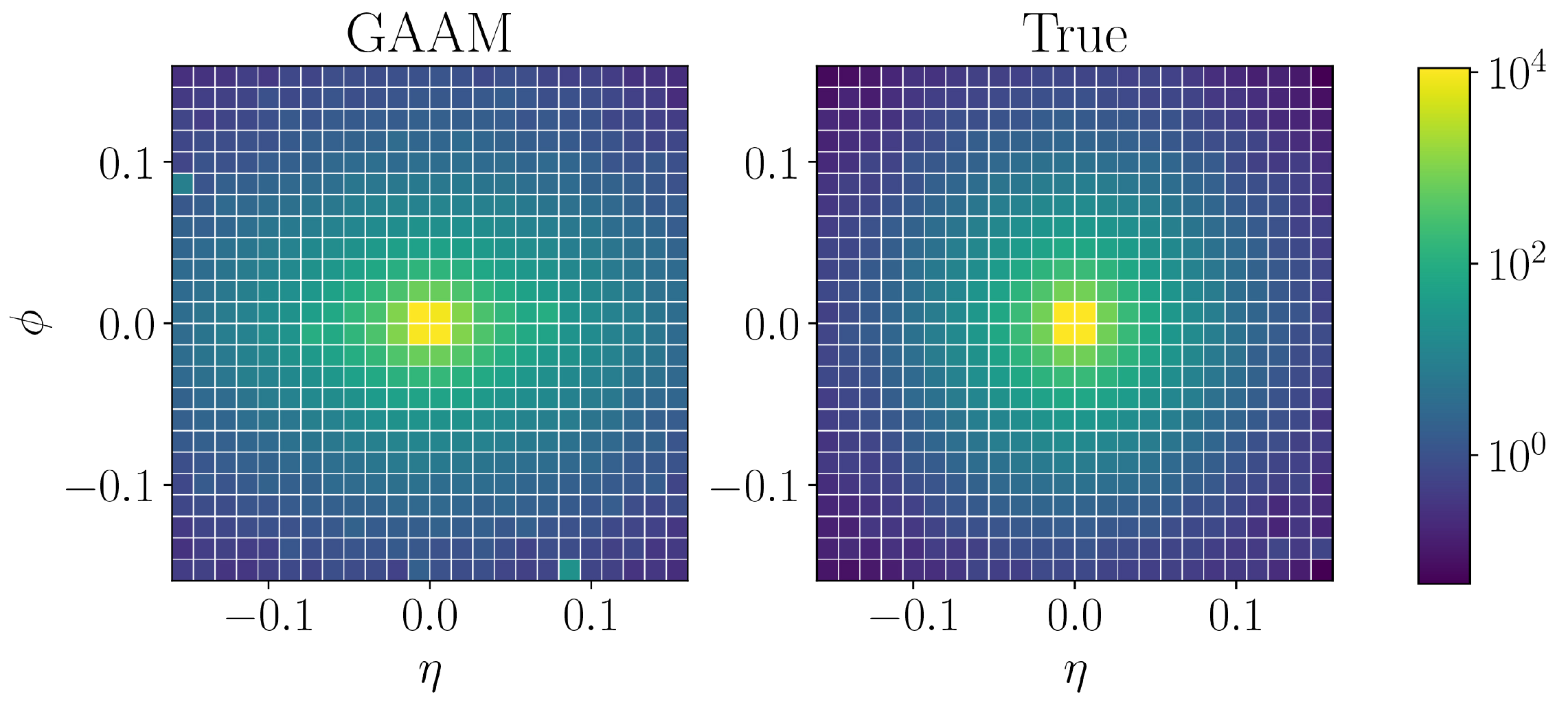}\label{meanimage_stripelayer_24x24}}\hfill
  \subfloat[Middle layer - (24, 12)]{\includegraphics[width=.5\textwidth]{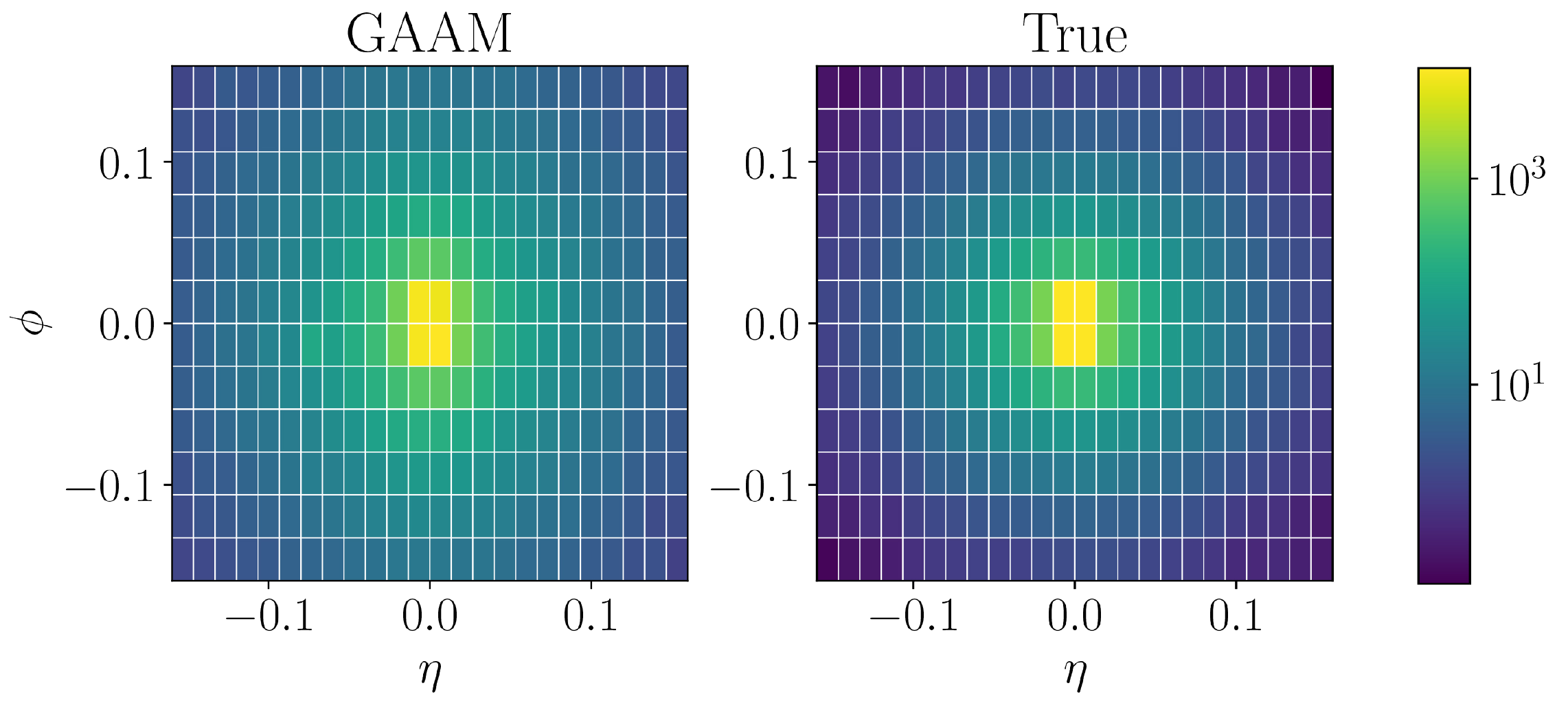}\label{meanimage_middlelayer_24x12}}\hfill
  \caption{ Average generated calorimeter images from GAAM for two geometries, (a) (24, 24) and (b) (24, 12), which do not appear in the training set and require interpolation. Each are compared to the Geant4 truth.}
  \label{new_shape_average_image}
\end{figure}

\begin{figure}[t!]
  \centering
  \subfloat[Middle layer - (24, 24)]{\includegraphics[width=.5\textwidth]{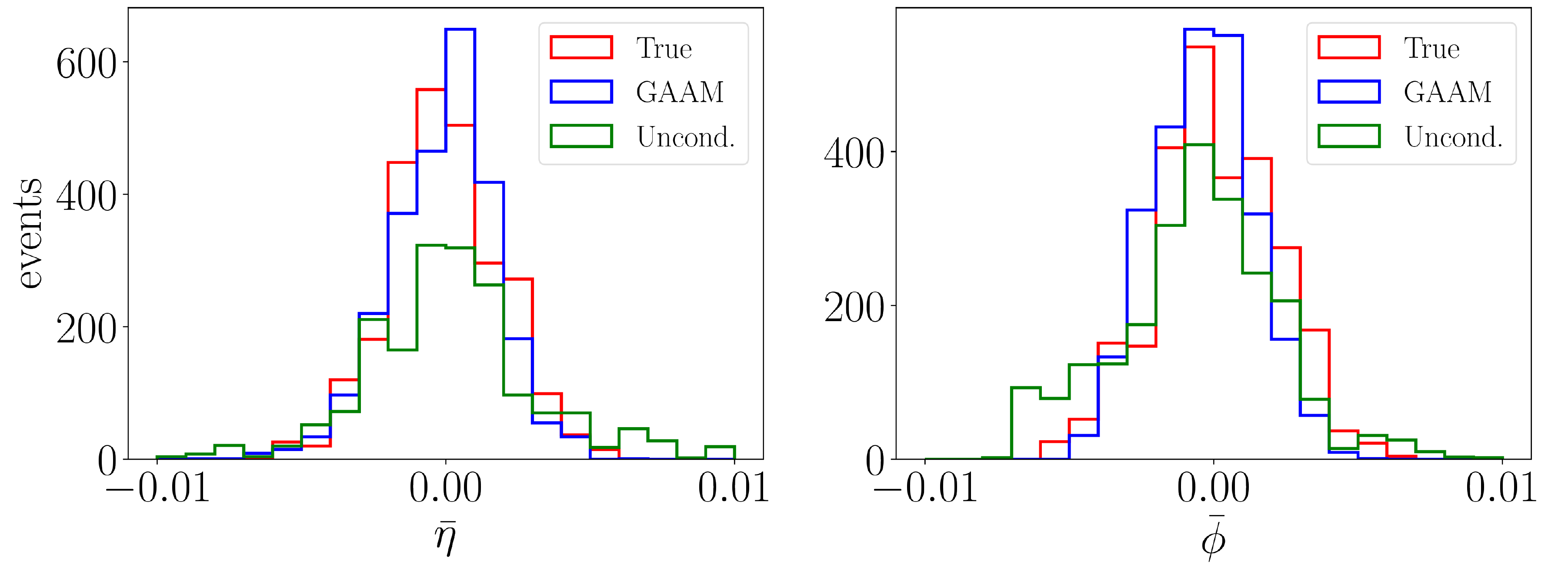}\label{eng-weighted_avgs_innerlayer_24x24}}\hfill
  \subfloat[Middle layer - (24, 12)]{\includegraphics[width=.5\textwidth]{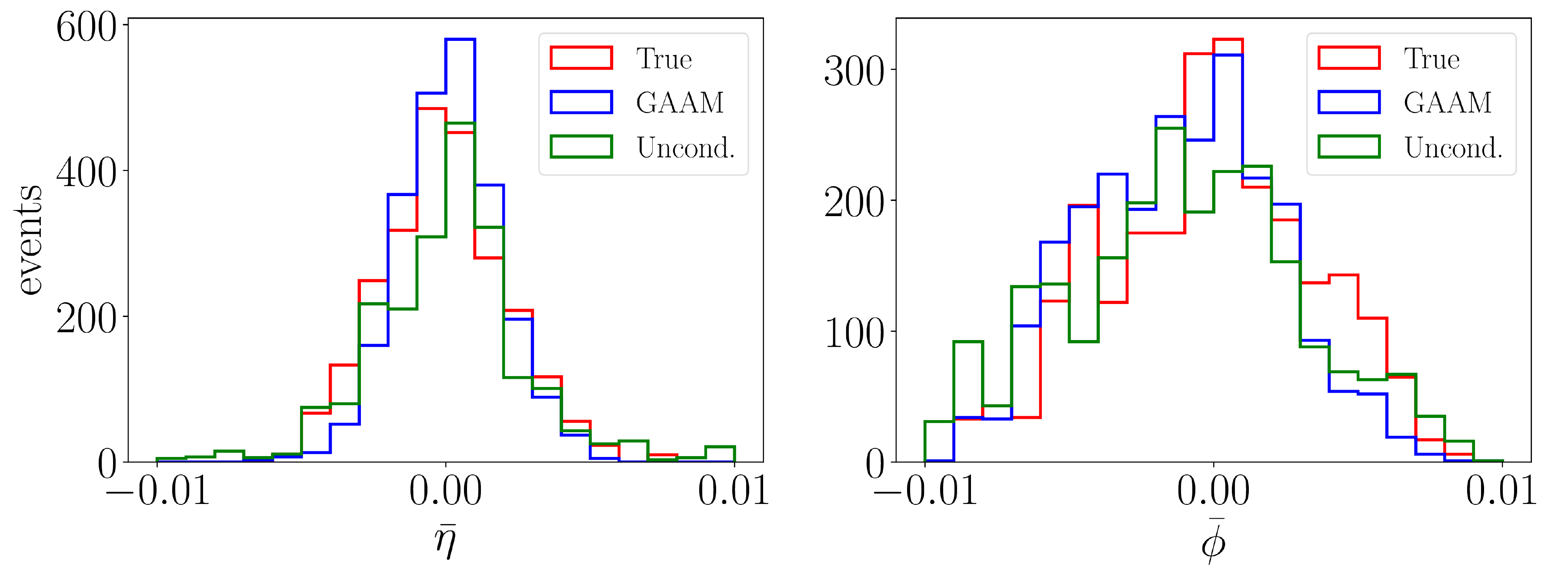}\label{eng-weighted_avgs_middlelayer_24x12}}\hfill\\
  \caption{     Distribution of the energy weighted means, $\bar \eta$ and $\bar \phi$, in samples generated by the GAAM, the unconditional model and  Geant4 (True), for several calorimeter layers with unseen varying segmentation in $(\eta,\ \phi)$, geometries which do not appear in the training set and require interpolation. The GAAM outperforms the unconditional baseline model consistently.}
  \label{new_shape_energy-weighted_avgs}
\end{figure}

\begin{figure}[h!]
  \centering
  \subfloat[Middle layer - (24, 24)]{\includegraphics[width=.5\textwidth]{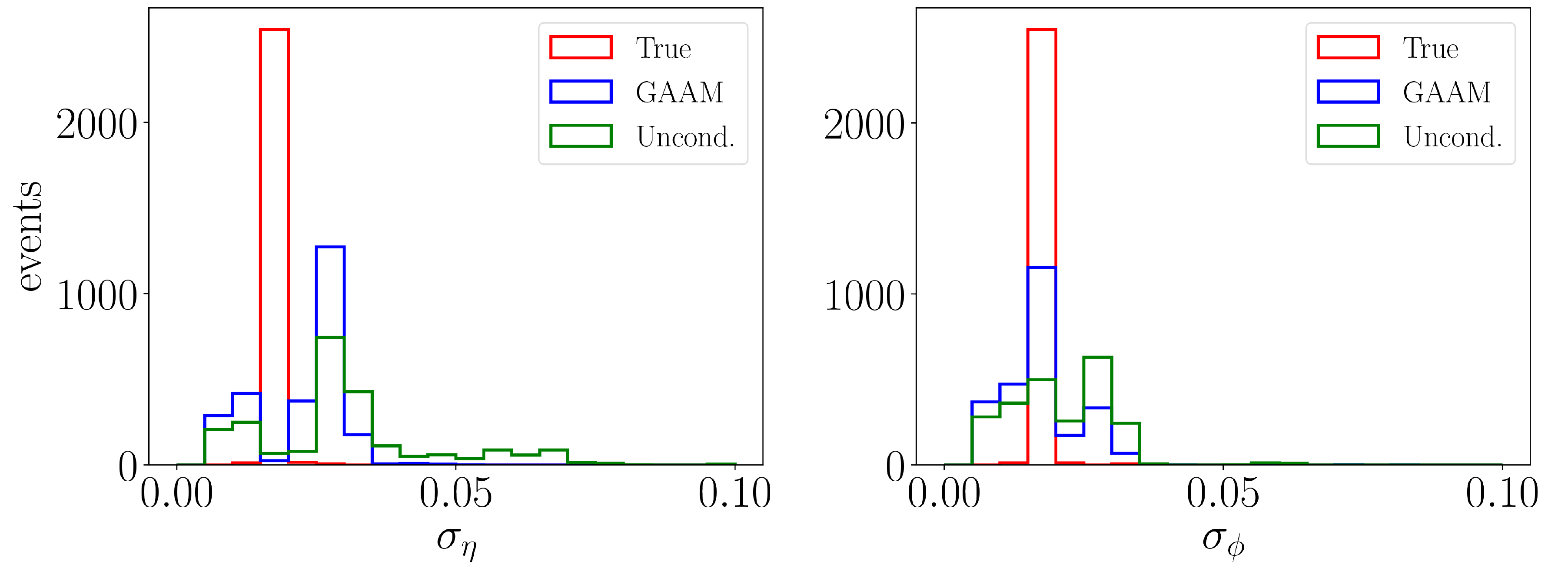}\label{shower_widths_middlelayer_inter_24x24}}\hfill
  \subfloat[Middle layer - (24, 12)]{\includegraphics[width=.5\textwidth]{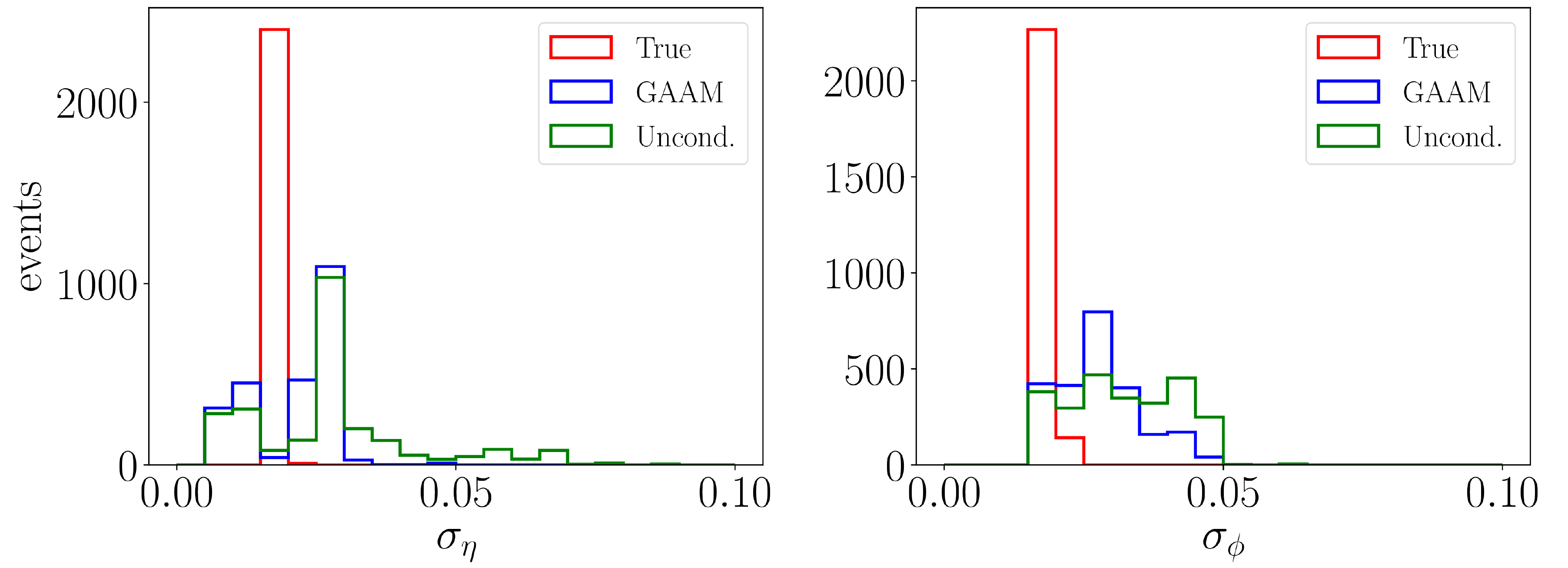}\label{shower_widths_middlelayer_inter_24x12}}\hfill\\
  \caption{Distribution of the shower widths, $\sigma_{\eta}$ and $\sigma_{\phi}$, in samples generated by the GAAM, the unconditional model and  Geant4 (True), for several calorimeter layers with unseen varying segmentation in $(\eta,\ \phi)$, geometries which do not appear in the training set and require interpolation. The GAAM outperforms the unconditional baseline model but it also struggles to produce narrow distributions.}
    \label{new_shape_shower_widths}
\end{figure}

Figure~\ref{new_shape_average_image} shows the average generated and true images for two unseen geometries which are bracketed by examples in the training set. Qualitatively, the GAAM has succeeded at interpolation.  A quantitative comparison is provided by examining the important physical quantities defined above; see Figs.~\ref{new_shape_energy-weighted_avgs}~and~\ref{new_shape_shower_widths} for histograms and Table~\ref{wdist_interp} for Wasserstein distances. The GAAM is able to reasonably interpolate to unseen geometries, although it struggles to produce very narrow shower widths.

\begin{figure}[t!]
  \centering
  \subfloat[Middle layer - (6, 6)]{\includegraphics[width=.5\textwidth]{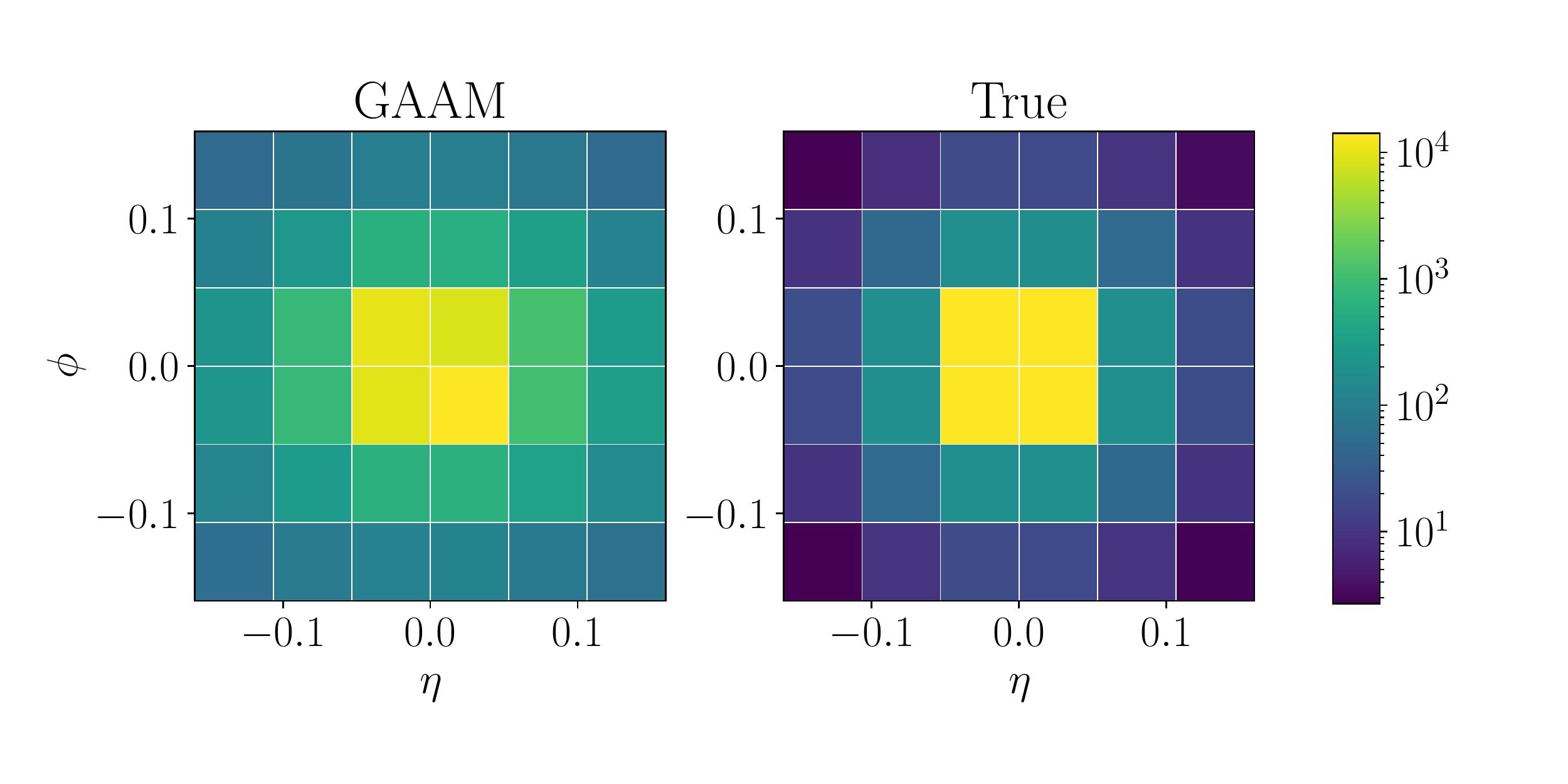}\label{meanimage_middlelayer_extra_6x6}}\hfill
  \subfloat[Middle layer - (96, 24)]{\includegraphics[width=.5\textwidth]{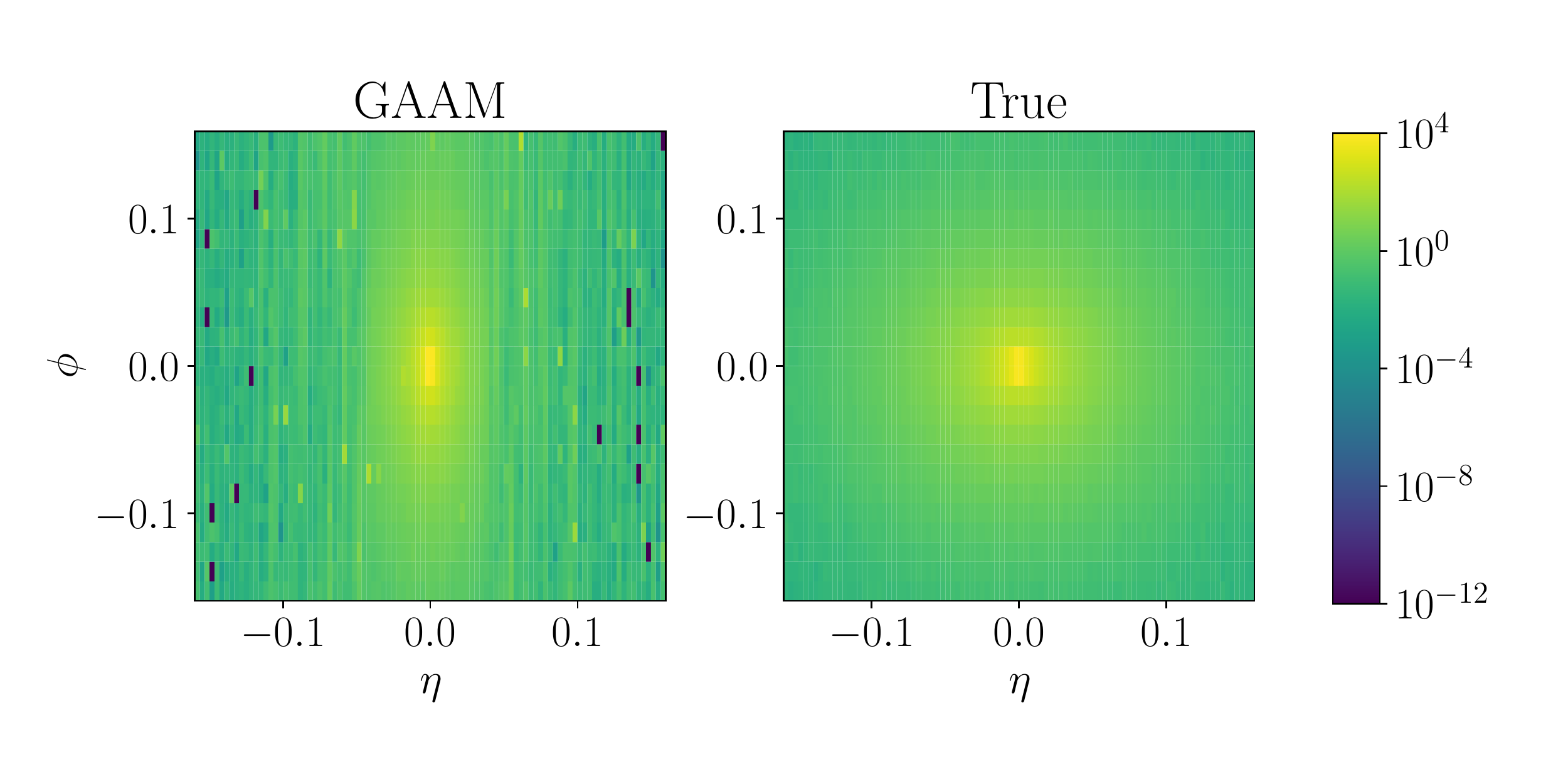}\label{meanimage_middlelayer_extra_96x24}}\hfill
  \hfill\\
  \subfloat[Middle layer - (9, 12)*]{\includegraphics[width=.5\textwidth]{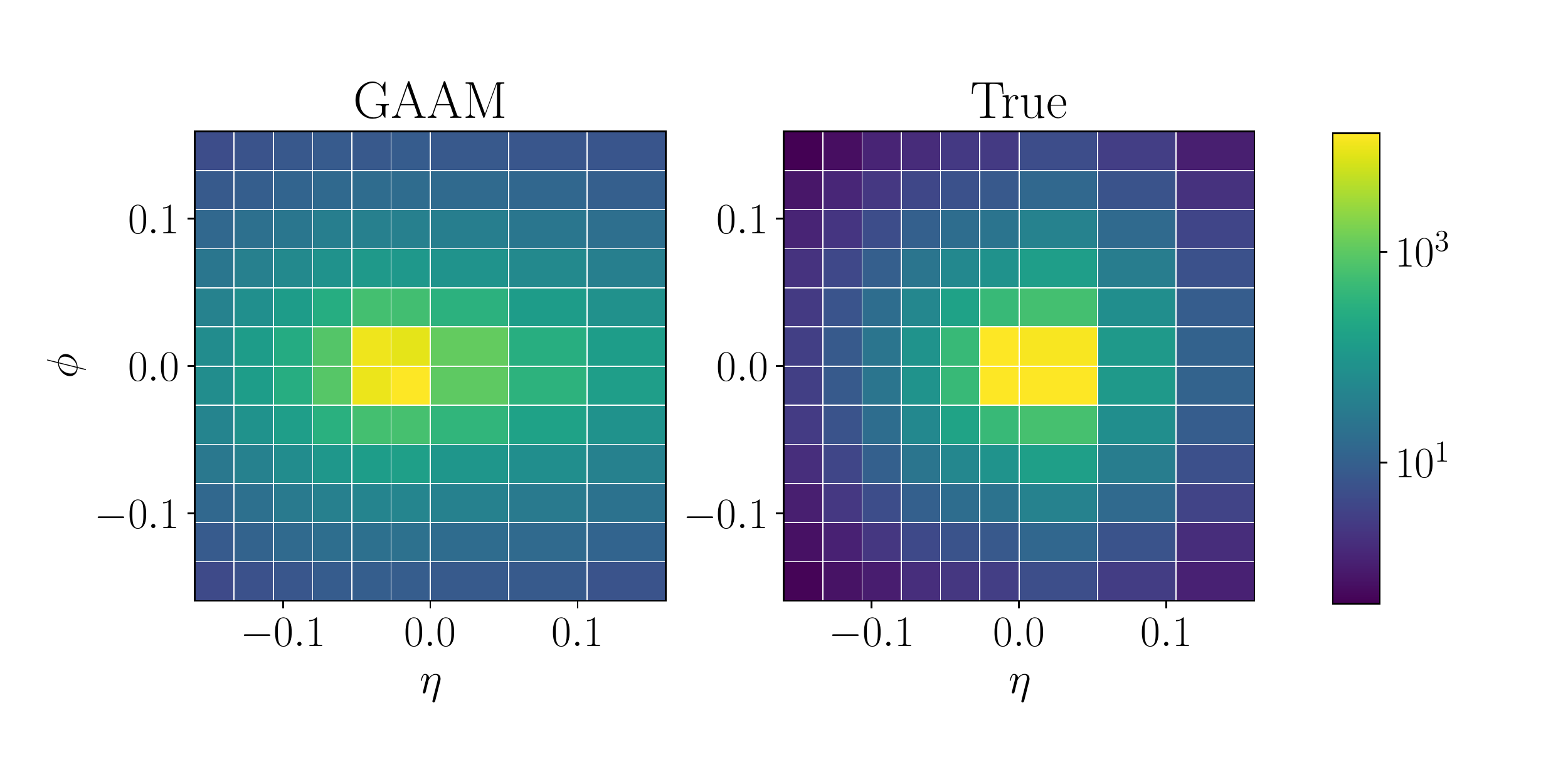}\label{meanimage_middlelayer_extra_9x12}}
  \caption{ Average generated calorimeter images from GAAM for two $(\eta,\ \phi)$ segmentations, (a) (6, 6), (b) (96, 24), and (c) (9, 12)*, which do not appear in the training set and require extrapolation. Each is compared to the Geant4 truth. GAAM does not extrapolate beyond the range of cell sizes seen in training.}
  \label{extra_shape_average_image}
\end{figure}

Extrapolation beyond the training examples given is much more challenging. We present three extrapolation geometries of the middle layer, which have cell sizes outside the range of the training data: (6, 6), (96, 24), and (9, 12)*. The (6, 6) geometry is composed of larger cell sizes than the largest cells in training data in both $\eta$-dimension and $\phi$-dimension. The (96, 24) geometry has a similar cell segmentation, but with cells half the size of the smallest cells in training data in the $\eta$-dimension and  twice the size of the smallest cells in training data in the $\phi$-dimension. Details about the spiral ordering in such cases are provided in Appendix~\ref{app:spiralOrder}. The (9, 12)* geometry is composed of two different cell sizes in the $\eta$-dimension, similar to the feature of boundary regions. The first six cells from the left are of the same size of the largest cells seen in training, and the other three cells are twice as large as the cells on the left. In the $\phi$-dimension, the cells are uniformly segmented. In contrast to interpolation, samples for unseen geometries which require extrapolation show significant differences between the generated and true images (Figure~\ref{extra_shape_average_image}),  with the GAAM failing to place the maximum energy at the centre of the image in the third case. For the (6, 6) geometry, the GAAM deposits insufficient energy in the central cells and too much in the exterior cells, and for (96, 24) too there are visible artifacts. Quantitative comparisons, not shown, confirm what is clear in the visualizations.

\section{Conclusion and Outlook}
\label{sec:Conclusion}

The Geometry-Aware Autoregressive Model is capable of learning how calorimeter cell response varies with the detector geometry, allowing it to rapidly generate simulated responses to a wide array of detector geometries, including those which do not appear in the training set.  This capacity for generalization eliminates the need to train and validate hundreds of models to describe the response of a typical calorimeter.  By conditioning the model on the properties of individual cells rather than the complete geometry~\cite{ATLAS:2022jhk}, the GAAM has the capacity to scale to many geometry configurations.




Further work is needed to improve the overall fidelity of the simulation, particularly for the high dimensional inner calorimeter layer. It may need to also be trained on data with a variable position of the incident particle and boundary regions that are not always at the centre of the image for real world applications. Additionally, one may compare such a geometry-aware approach with a \textit{geometry-agnostic} approach to shower simulation with point-cloud models~\cite{Kansal:2021cqp,Kansal:2022spb,Leigh:2023toe,Mikuni:2023dvk} to determine the suitability of each approach on a range of detector geometries. While the geometries investigated were motivated by the ATLAS calorimeter, further work would be required to generalize to even more diverse geometries such as hexagonal cells or varying number of calorimeter layers relevant to other calorimeters such as in CMS. While not studied in this work, GAAM could be trained on non-uniform segmentations in the $\phi$ direction if needed. As the first attempt at designing generative models that can handle variable calorimeter geometries, this work did not focus on optimization of the speed of the GAAM, which is left for future work. As with previous work such as CaloFlow~\cite{Krause:2021ilc} and the faster CaloFlow~II~\cite{Krause:2021wez} or caloScore~\cite{Mikuni:2022xry} and the faster caloScore~v2\cite{Mikuni:2023tqg}, future work can be expected to speed up geometry-aware models, for example with the use of ML-based dimensionality reduction methods.

This study is also the first step towards building `foundational models' in HEP, general purpose machine learning models trained on large and diverse data, intended to be fine-tuned to individual tasks. The GAAM demonstrates the feasibility of general purpose generative models for calorimeter simulation. These would act as foundational models, trained on various calorimeters and geometries, and then shared between experiments. The experiments would only need to fine-tune the model to their specific calorimeter, requiring far less training data, computing resources and human resources compared to developing a generative model from scratch for their specific calorimeter.

\acknowledgments

DW, AG and DS are supported by The Department of Energy Office of Science grant DE-SC0009920. AG is also supported under contract DE-AC02-05CH11231, and DS is also supported by the HEPCAT Fellowship under contract DE-SC0022313. We are grateful to Benjamin Nachman for help with the Geant4 setup. We also thank Tobias Golling and Johnny Raine for fruitful discussions.

\paragraph{Note added:} As this manuscript was being finalized, we became aware of the recent work in Ref.~\cite{Buhmann:2023bwk}, which uses point-cloud generative model to simulate the future International Large Detector (ILD) calorimeter. It does not make comparisons on multiple challenging geometries as we do.


\bibliographystyle{JHEP}
\bibliography{main.bib}

\clearpage
\appendix
\begin{center}
    \LARGE{Appendix}
\end{center}

\section{Spiral Ordering}
\label{app:spiralOrder}
\begin{figure}[h!]
      \centering
      \includegraphics[width=.5\textwidth]{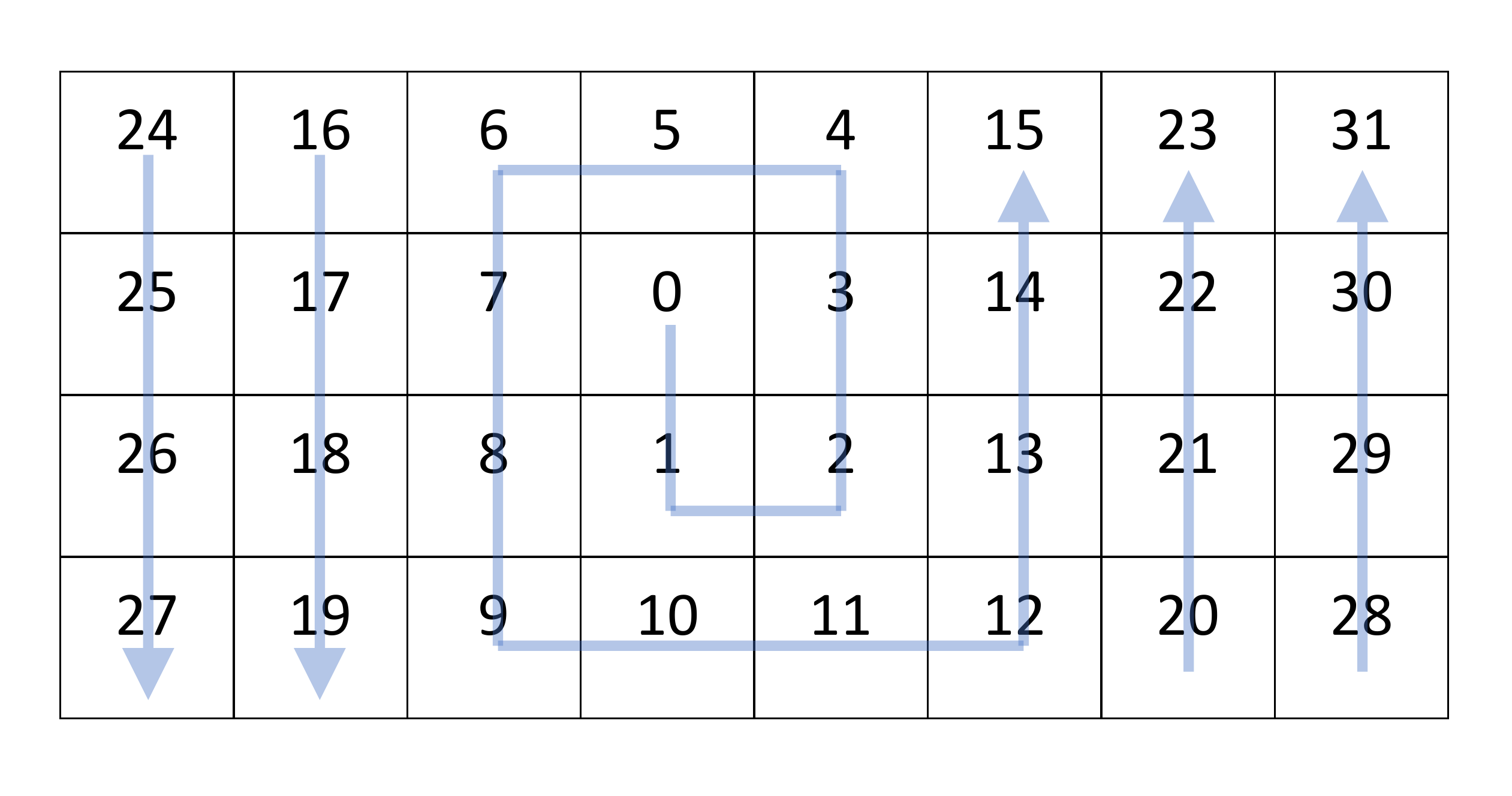}\label{}
      \captionsetup{width=.75\textwidth}\caption{Illustration of how the 2D image is flattened in a spiral order when the image is non-square.}
      \label{non-square}
  \end{figure}

There are special cases where the geometry does not fit perfectly into our data preprocessing approach. For example, when the events have non-square geometries, such as in the inner layer, we have a modified version of the spiral path shown in Figure~\ref{non-square}. The spiral path will look the same for the center square until it reaches the boundary in one dimension. Then, the spiral path will go left and right in turn to cover the peripheral cells.

In another case, with an even number of cells in $\eta$ and $\phi$ direction, there are four cells closest to the position of the shower center. We choose the top-left cell in the central four cells. The known prior distribution is obtained by sampling from the training data.

\section{Individual Shower Images}
\begin{figure}[h!]
  \centering
  \subfloat[Inner layer - (48, 4)]{\includegraphics[width=.5\textwidth]{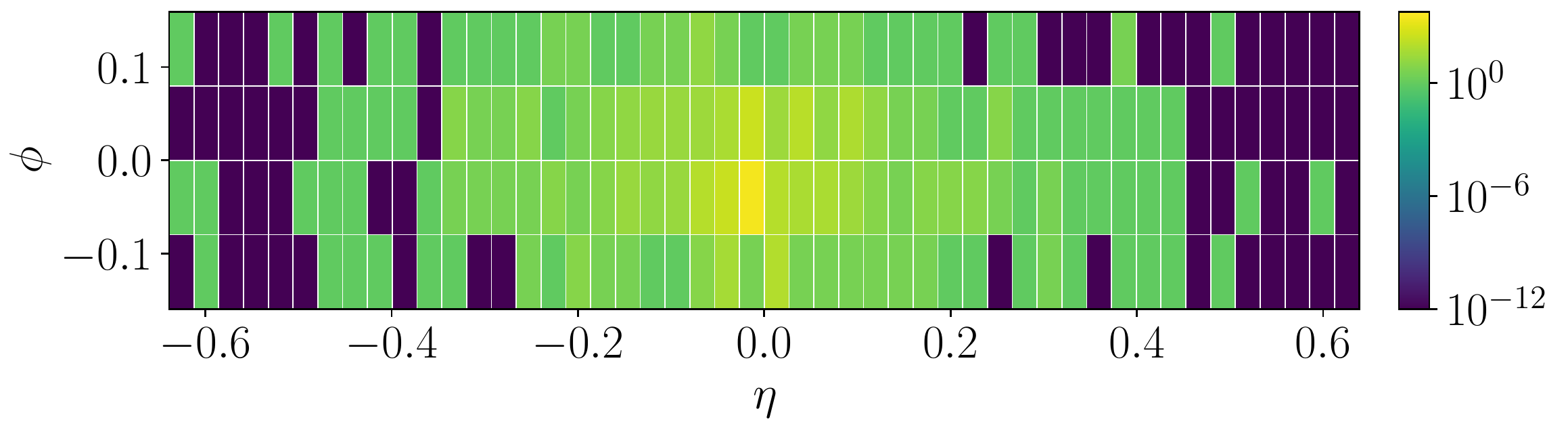}\label{}}\hfill
  \subfloat[Inner layer - (48, 12)]{\includegraphics[width=.5\textwidth]{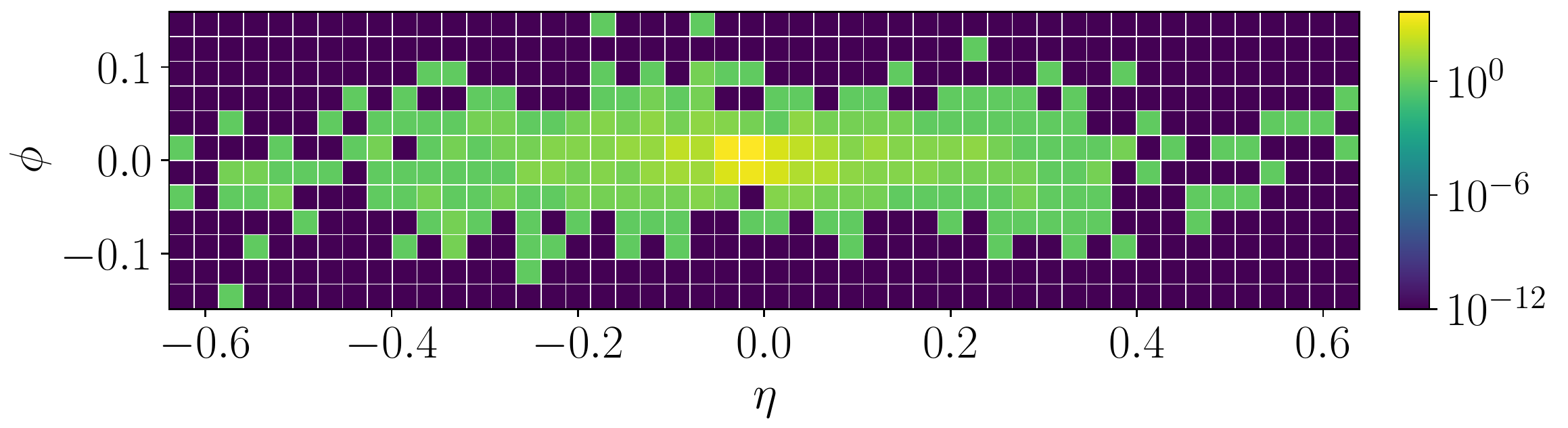}\label{}}\hfill
  \subfloat[Inner layer - (192, 4)]{\includegraphics[width=.5\textwidth]{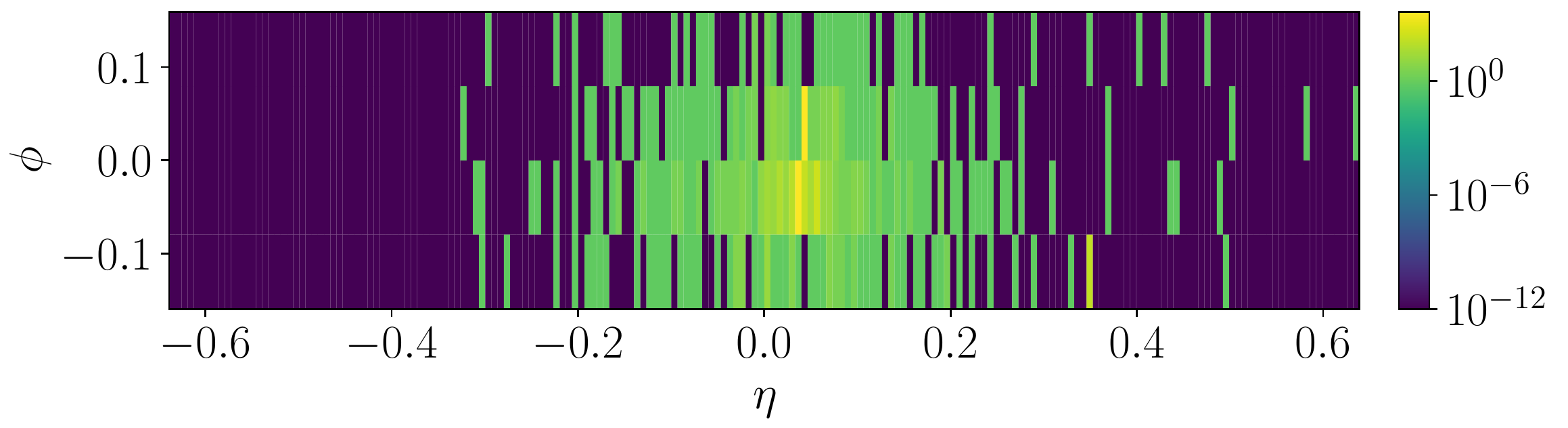}\label{}}\hfill
  \subfloat[Inner layer - (192, 48)]{\includegraphics[width=.49\textwidth]{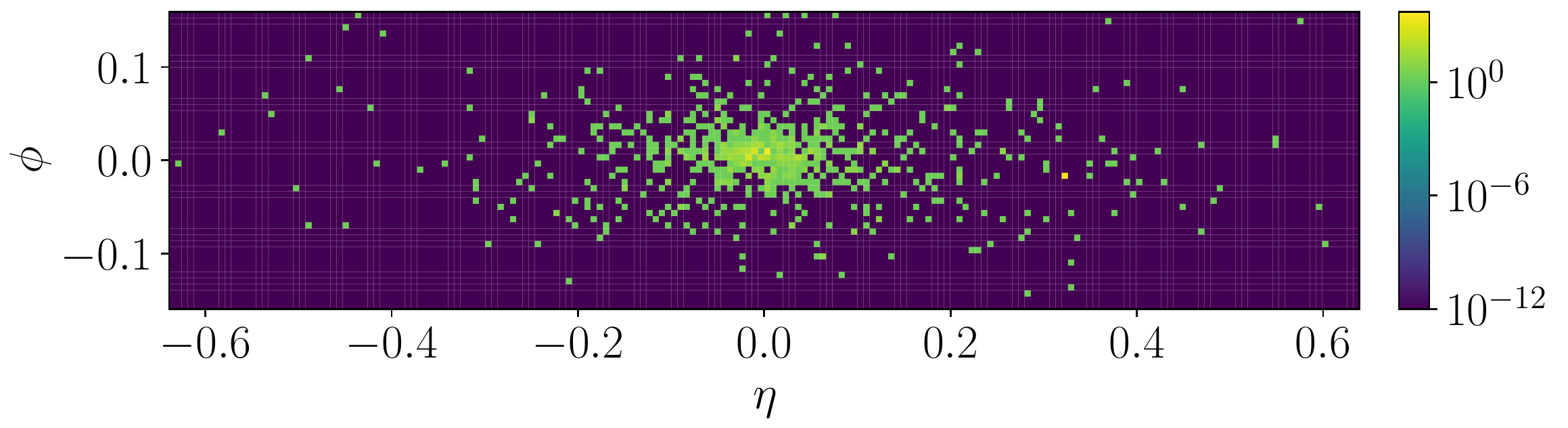}\label{}}\hfill
  \caption{Examples of individual events of the inner layer generated by GAAM}
  \label{gaam_individual_inner}
\end{figure}

\begin{figure}[h!]
  \centering
  \subfloat[Inner layer - (48, 4)]{\includegraphics[width=.5\textwidth]{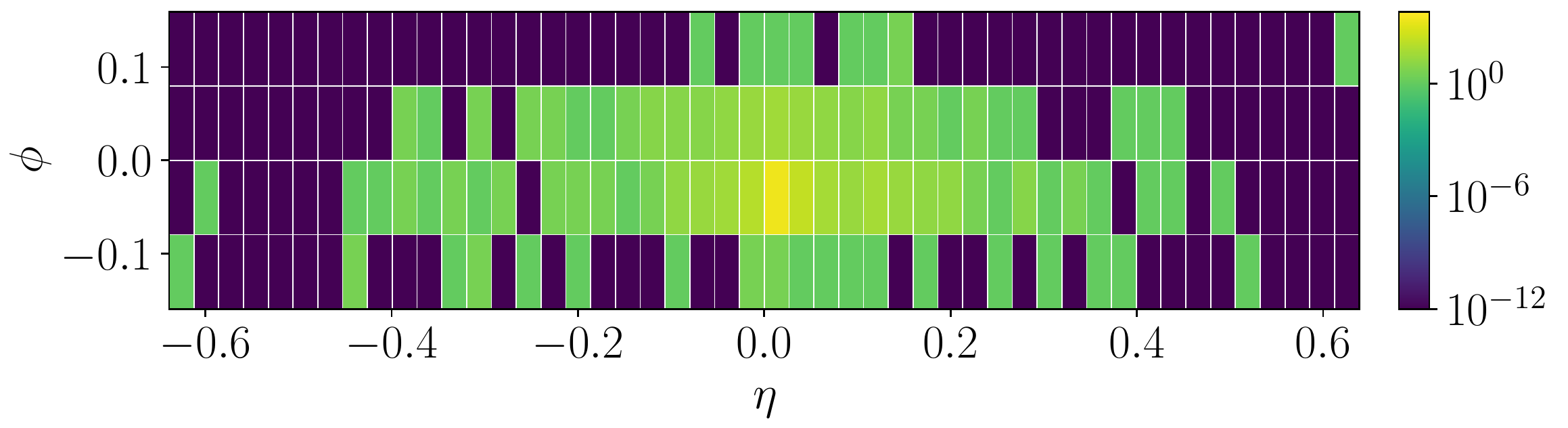}\label{}}\hfill
  \subfloat[Inner layer - (48, 12)]{\includegraphics[width=.5\textwidth]{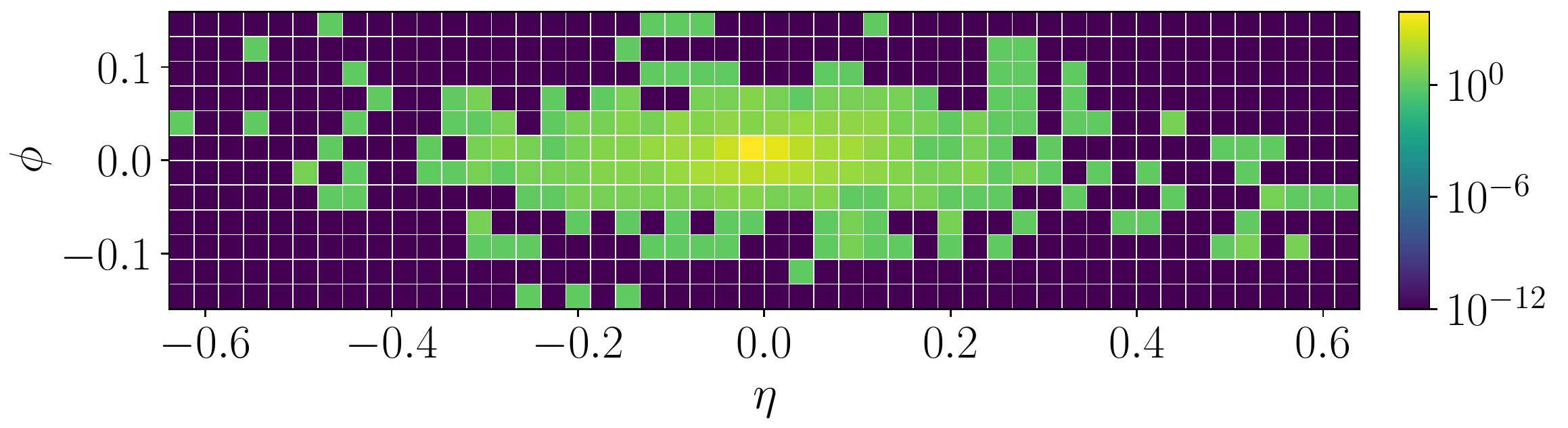}\label{}}\hfill
  \subfloat[Inner layer - (192, 4)]{\includegraphics[width=.5\textwidth]{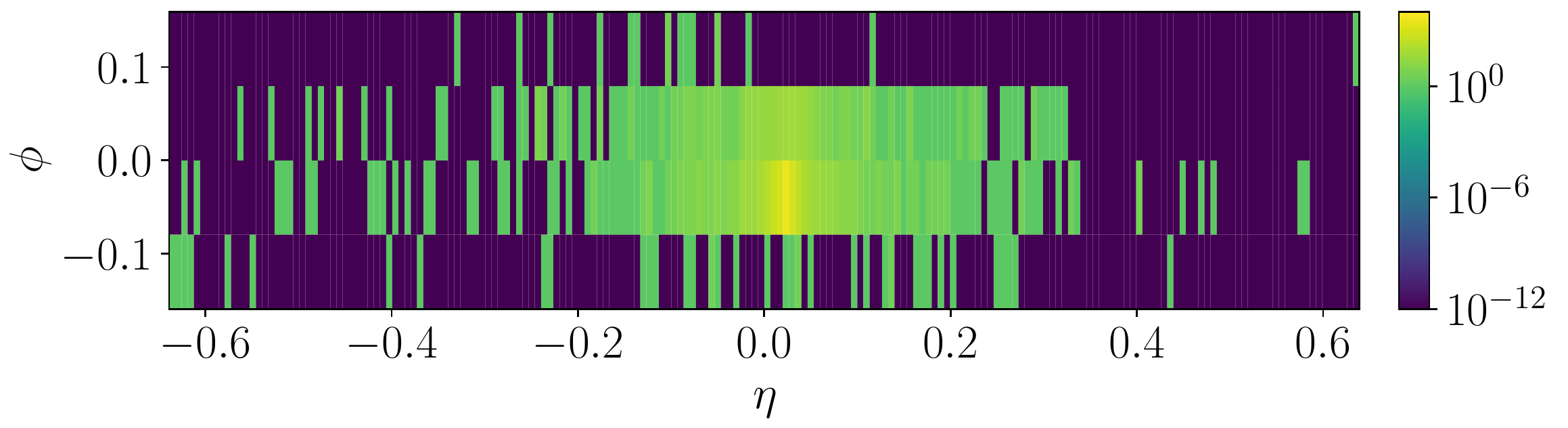}\label{}}\hfill
  \subfloat[Inner layer - (192, 48)]{\includegraphics[width=.49\textwidth]{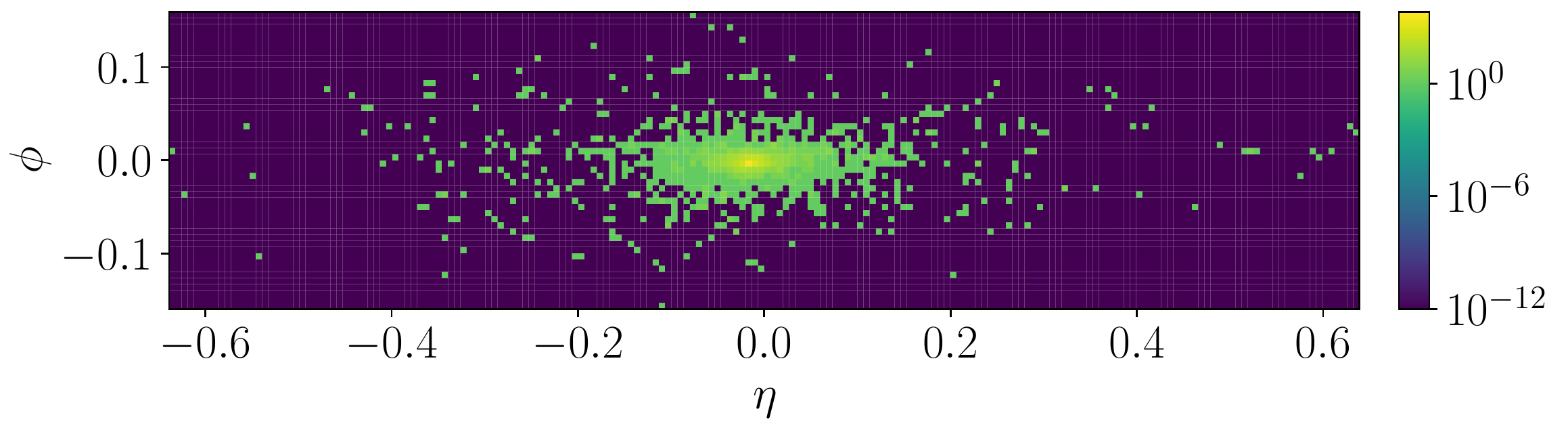}\label{}}\hfill
  \caption{Examples of individual events of the inner layer generated by Geant4}
  \label{true_individual_inner}
\end{figure}

\begin{figure}[h!]
  \centering
  \subfloat[Middle layer - (12, 12)]{\includegraphics[width=.5\textwidth]{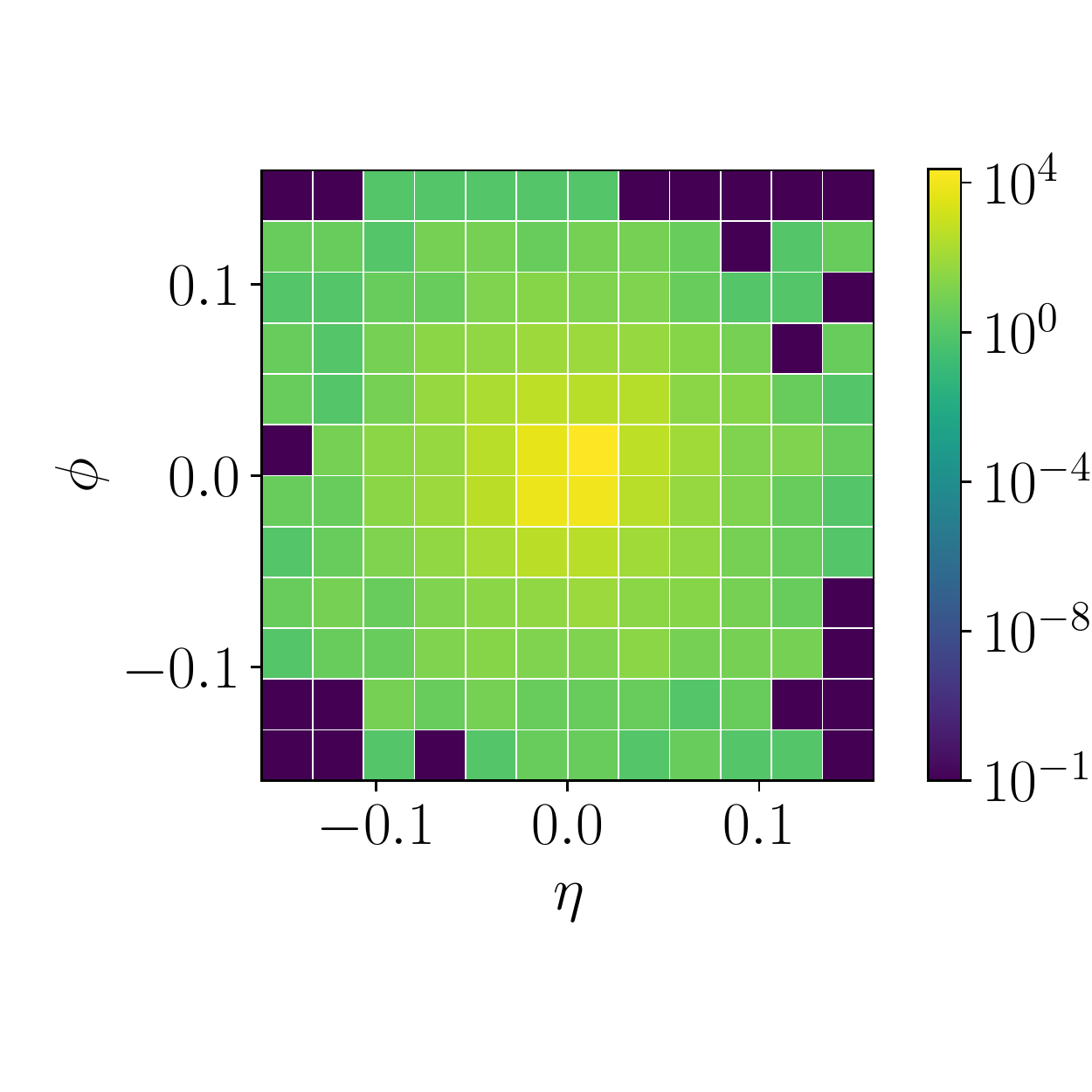}\label{}}\hfill
  \subfloat[Middle layer - (48, 24)]{\includegraphics[width=.5\textwidth]{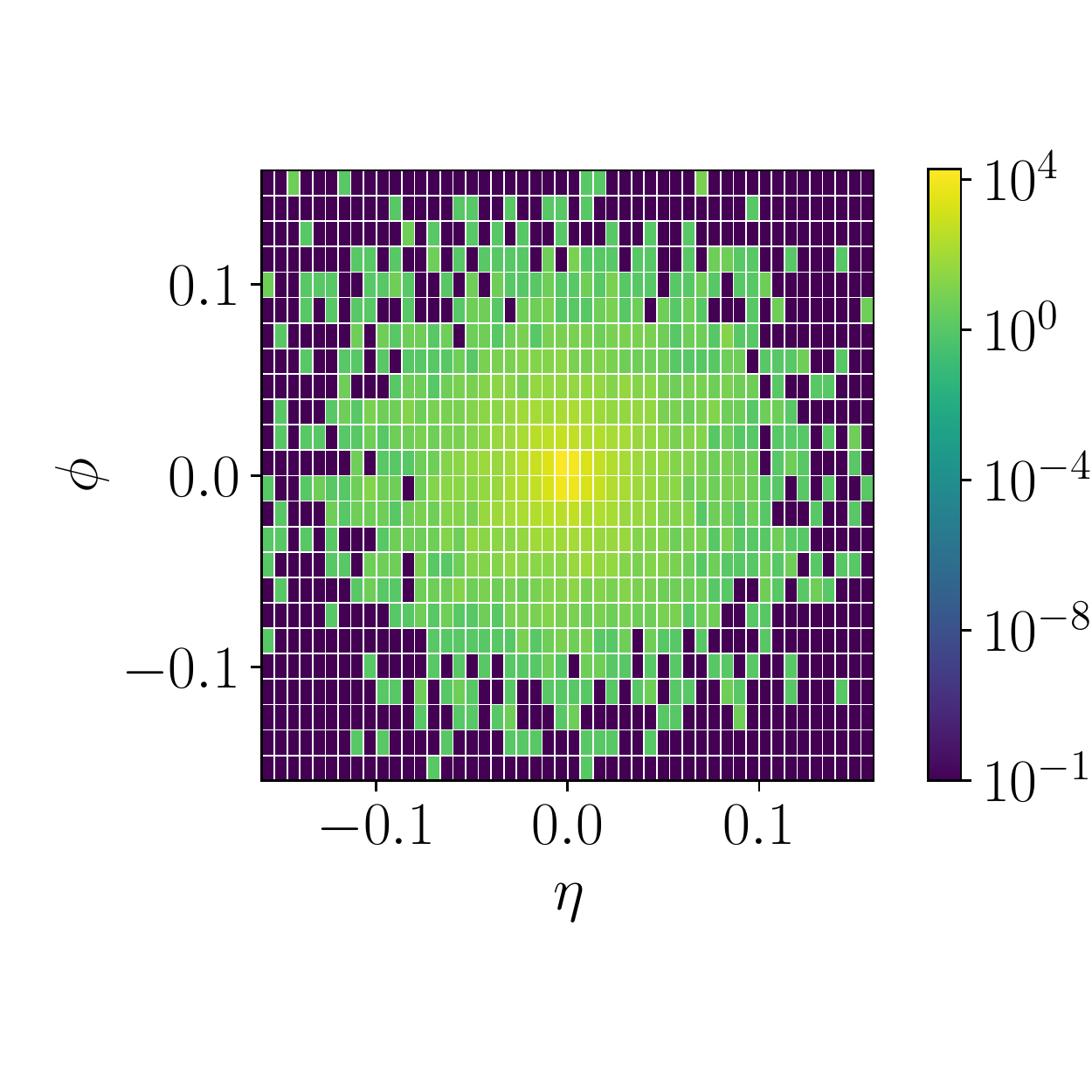}\label{}}\hfill
  \subfloat[Middle layer - (48, 48)]{\includegraphics[width=.5\textwidth]{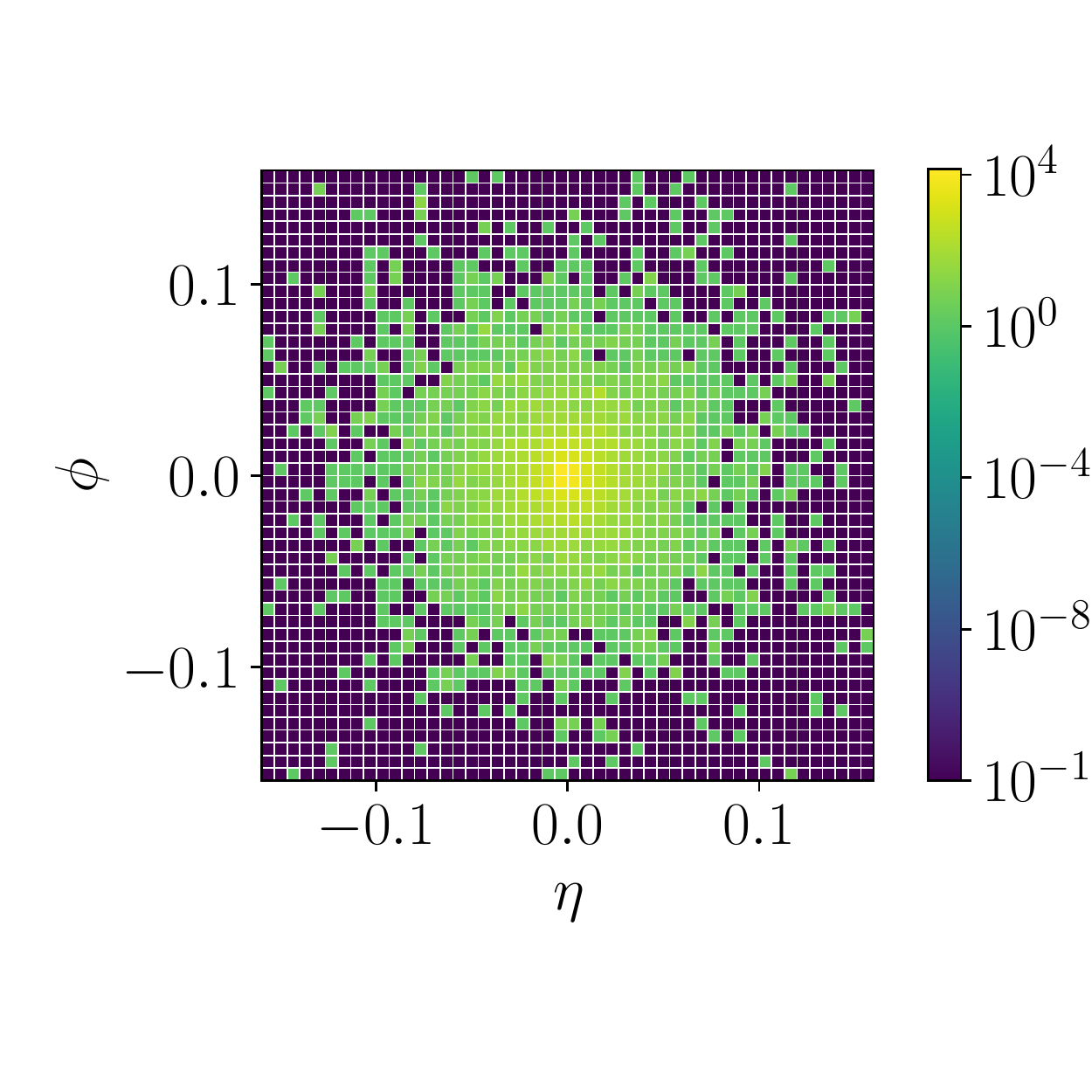}\label{}}\hfill
  \subfloat[Middle layer - (36, 48)*]{\includegraphics[width=.49\textwidth]{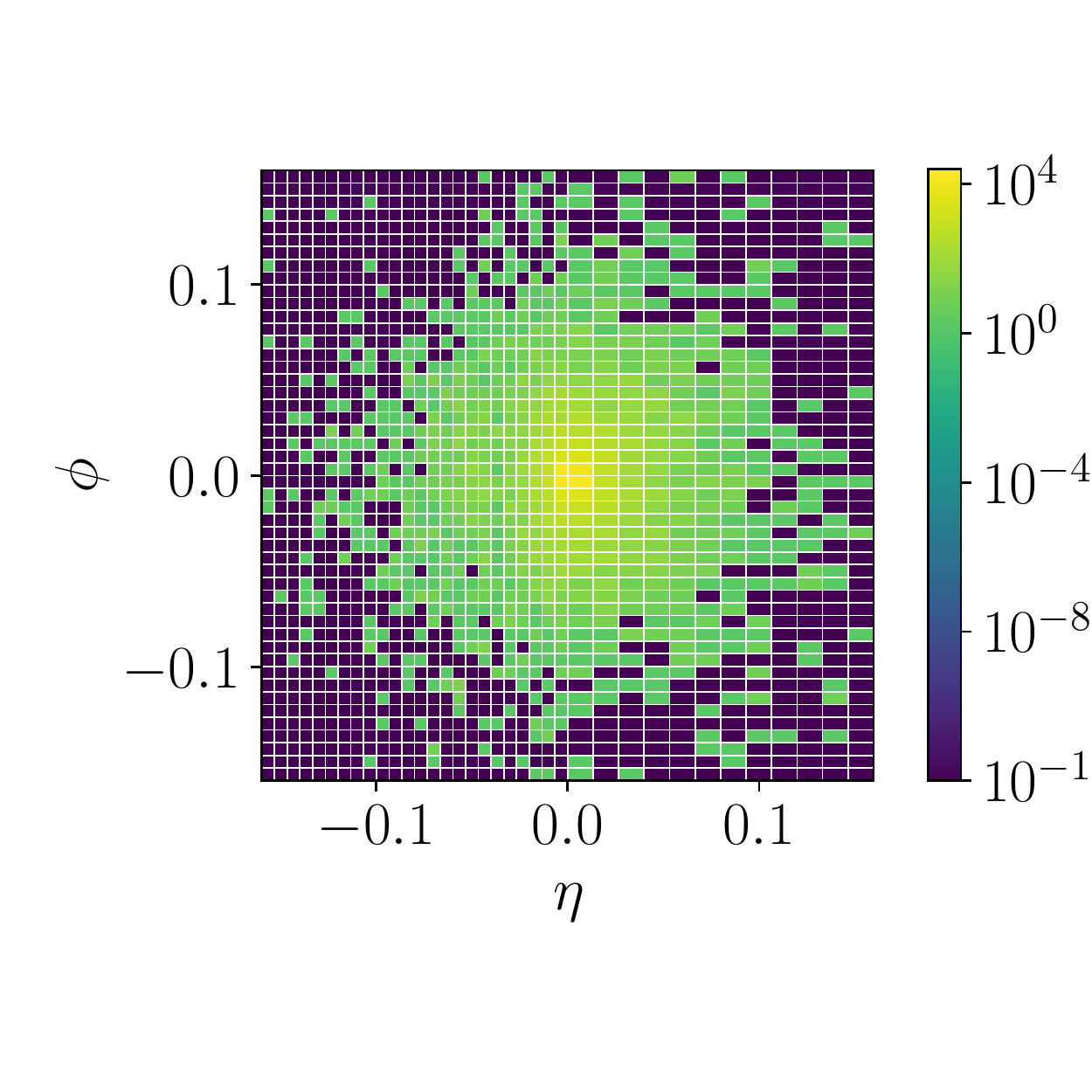}\label{}}\hfill
  \caption{Examples of individual events of the middle layer generated by GAAM}
  \label{gaam_individual_middle}
\end{figure}

\begin{figure}[h!]
  \centering
  \subfloat[Middle layer - (12, 12)]{\includegraphics[width=.5\textwidth]{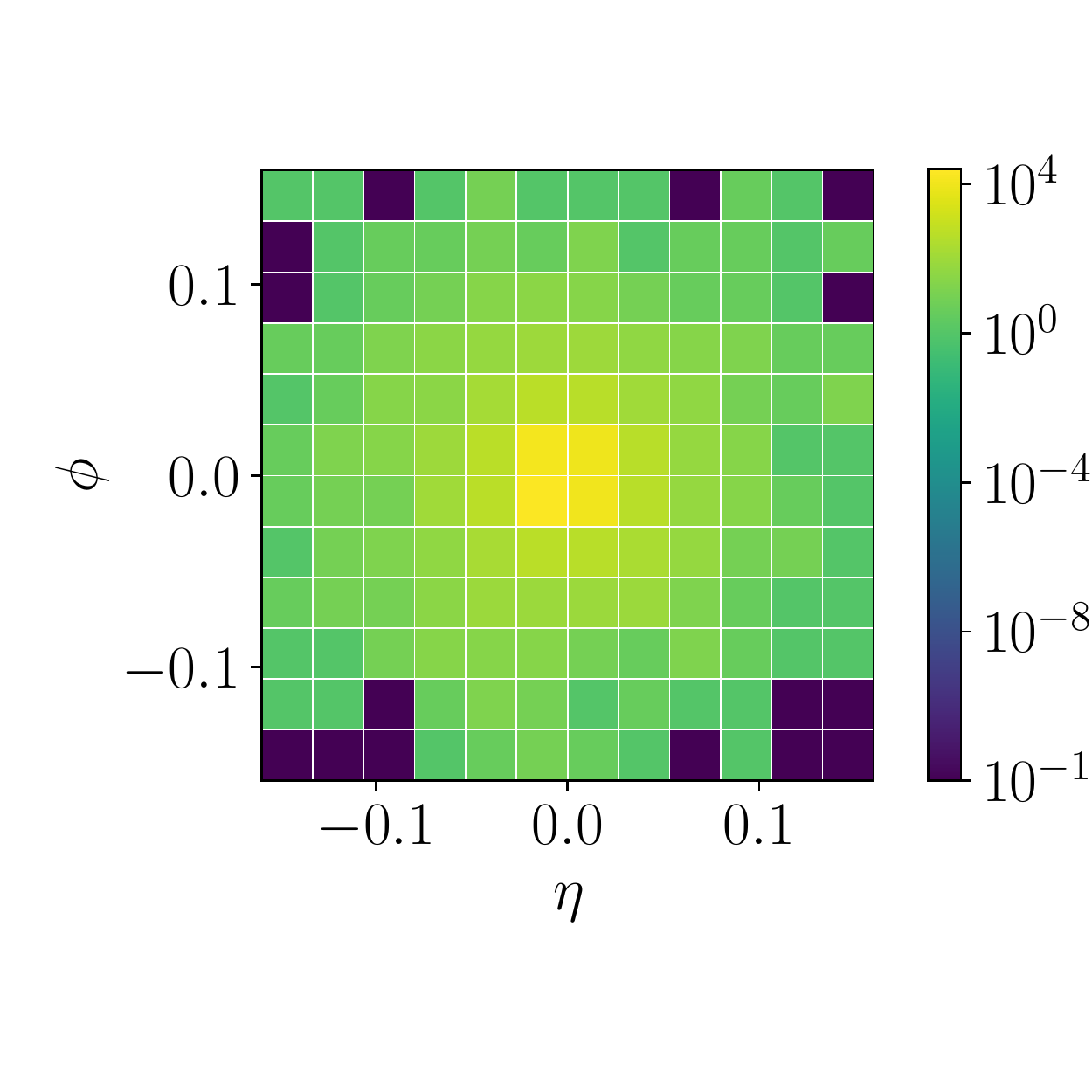}\label{}}\hfill
  \subfloat[Middle layer - (48, 24)]{\includegraphics[width=.5\textwidth]{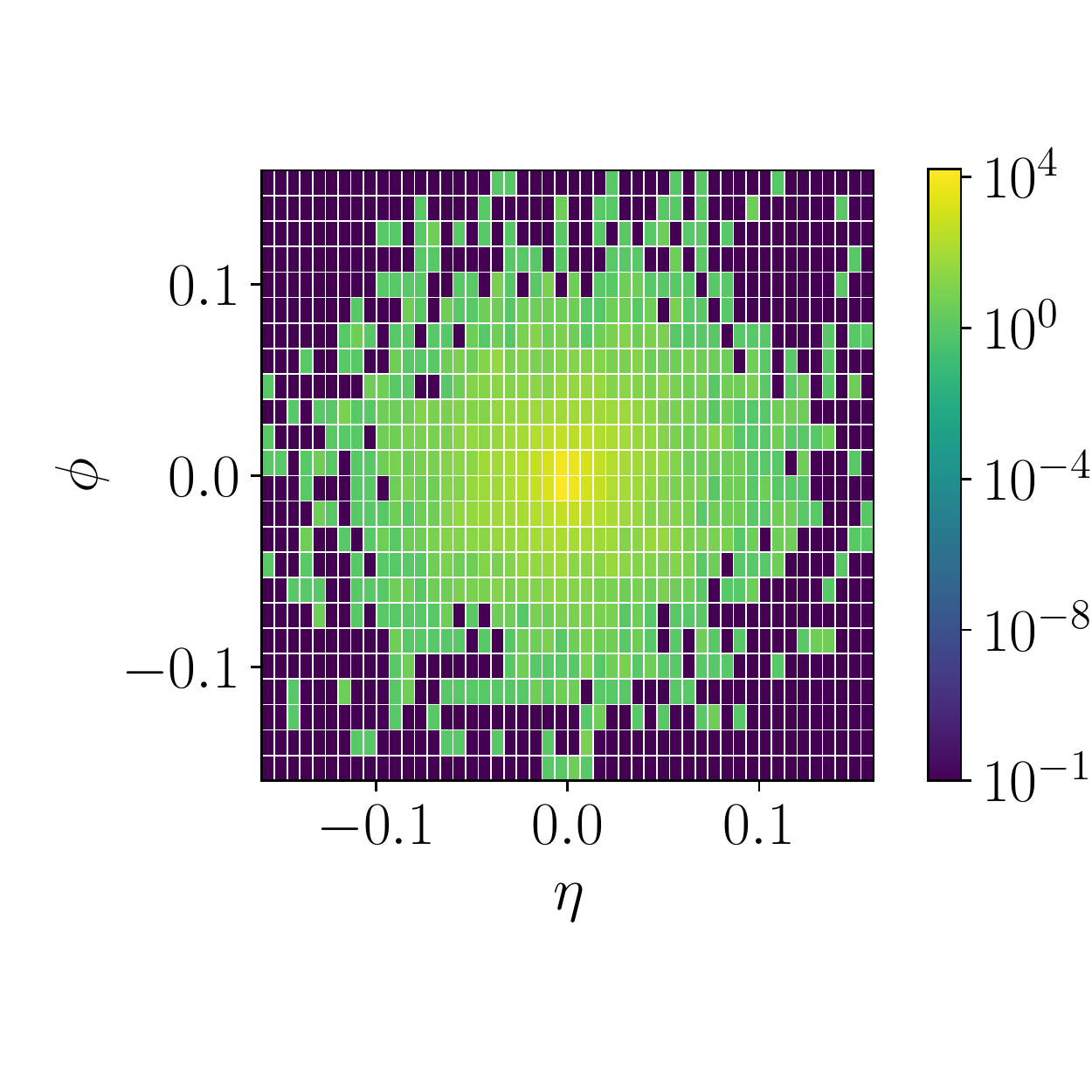}\label{}}\hfill
  \subfloat[Middle layer - (48, 48)]{\includegraphics[width=.5\textwidth]{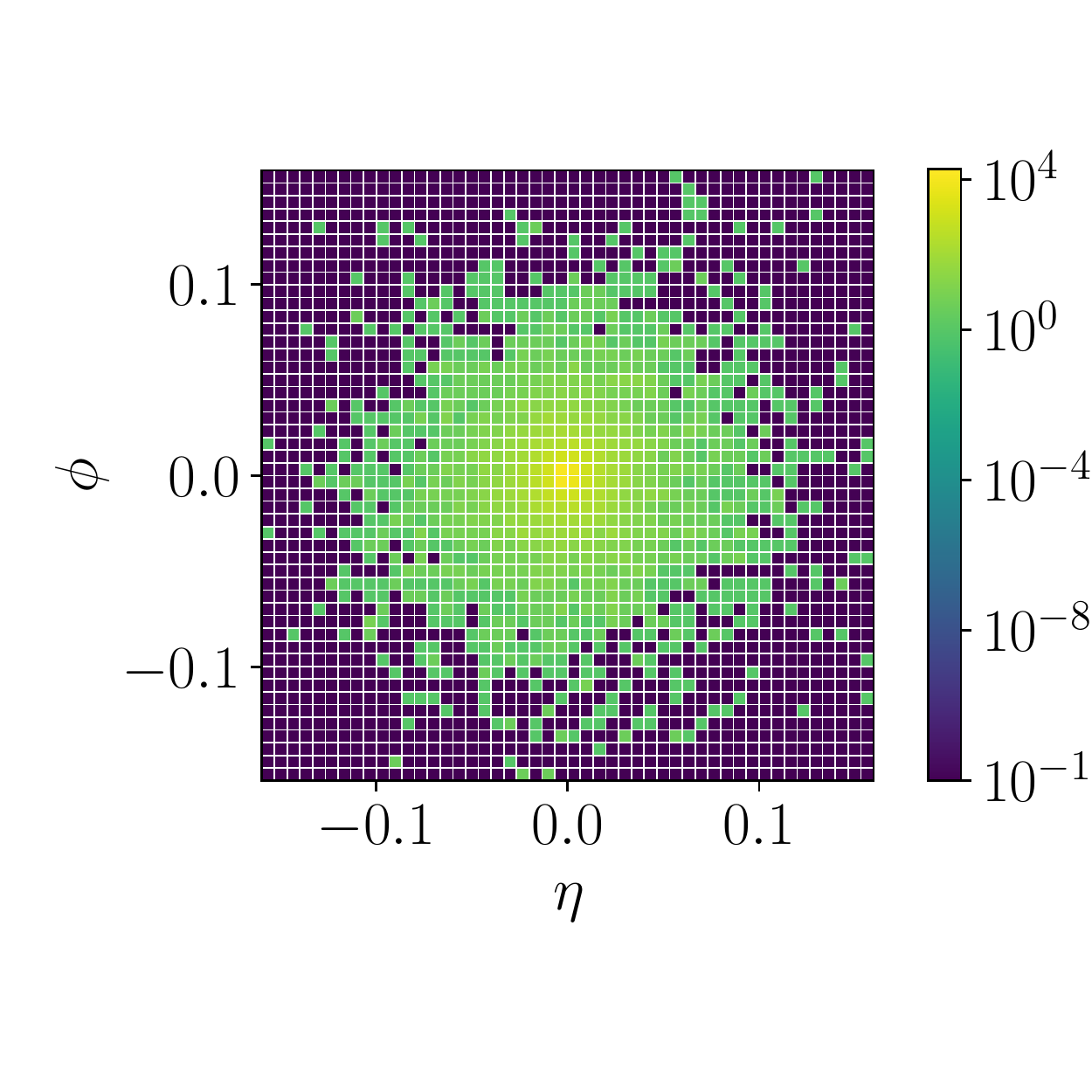}\label{}}\hfill
  \subfloat[Middle layer - (36, 48)*]{\includegraphics[width=.49\textwidth]{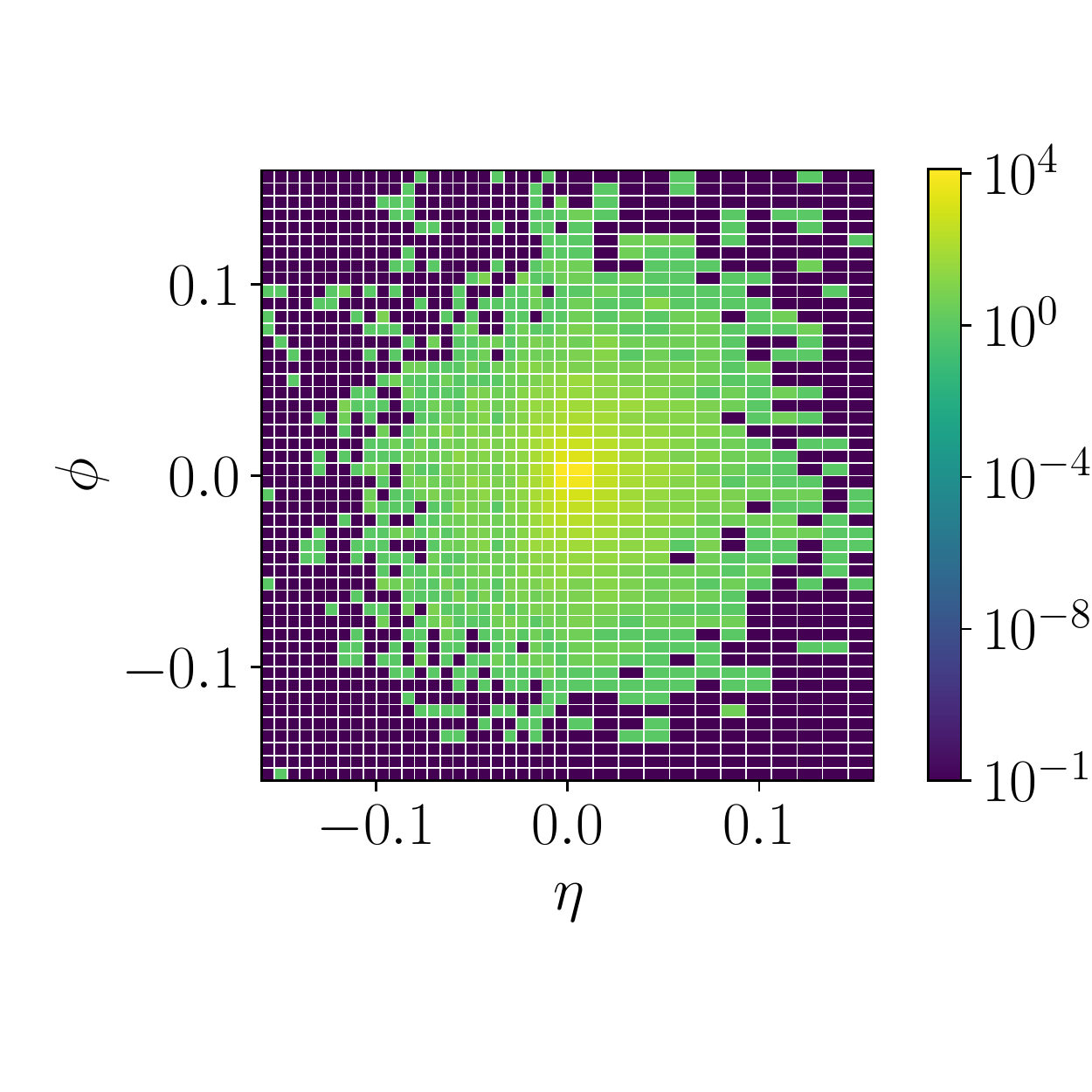}\label{}}\hfill
  \caption{Examples of individual events of the middle layer generated by Geant4}
  \label{true_individual_middle}
\end{figure}

\begin{figure}[h!]
  \centering
  \subfloat[Outer layer - (24, 24)]{\includegraphics[width=.5\textwidth]{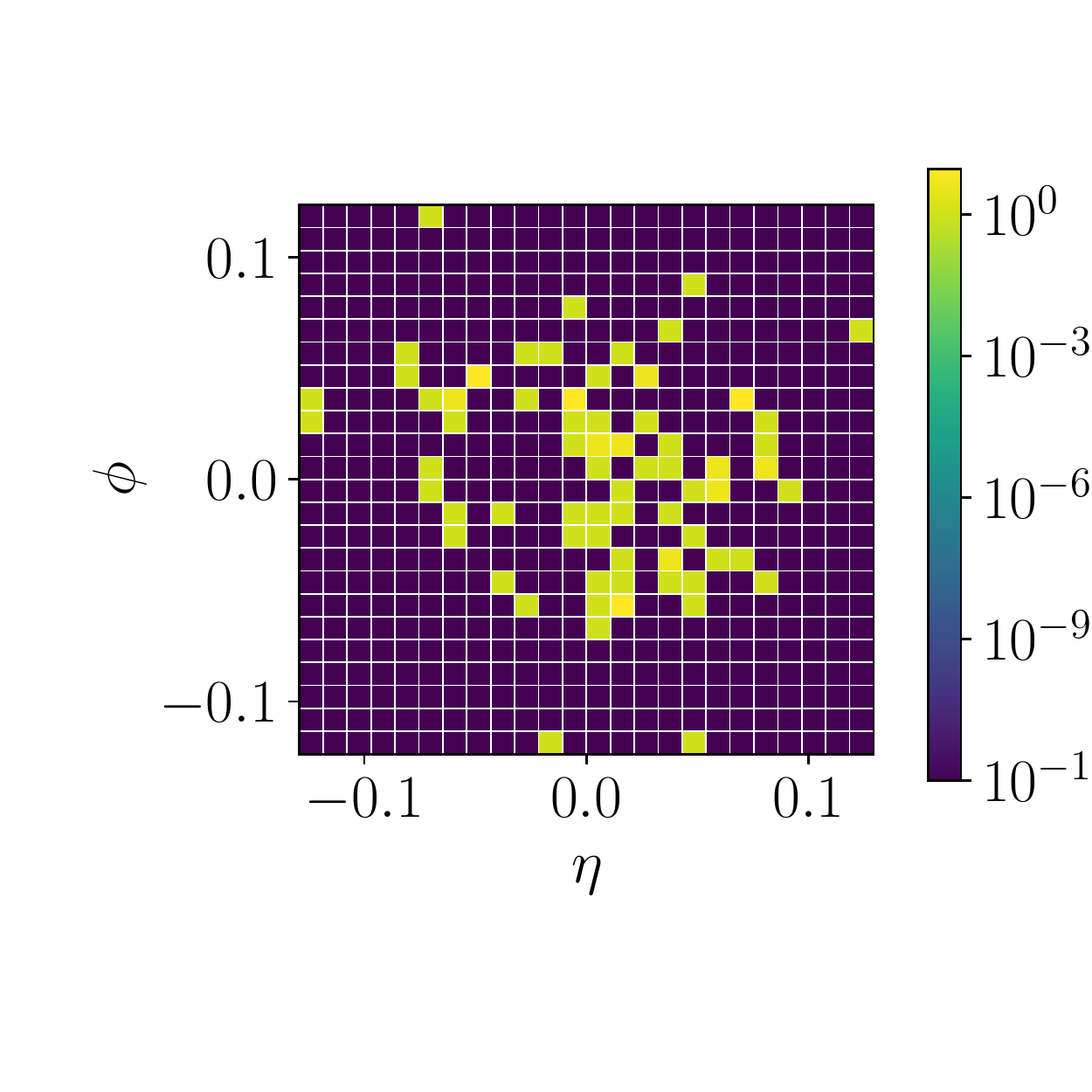}\label{}}\hfill
  \subfloat[Outer layer - (24, 24)]{\includegraphics[width=.5\textwidth]{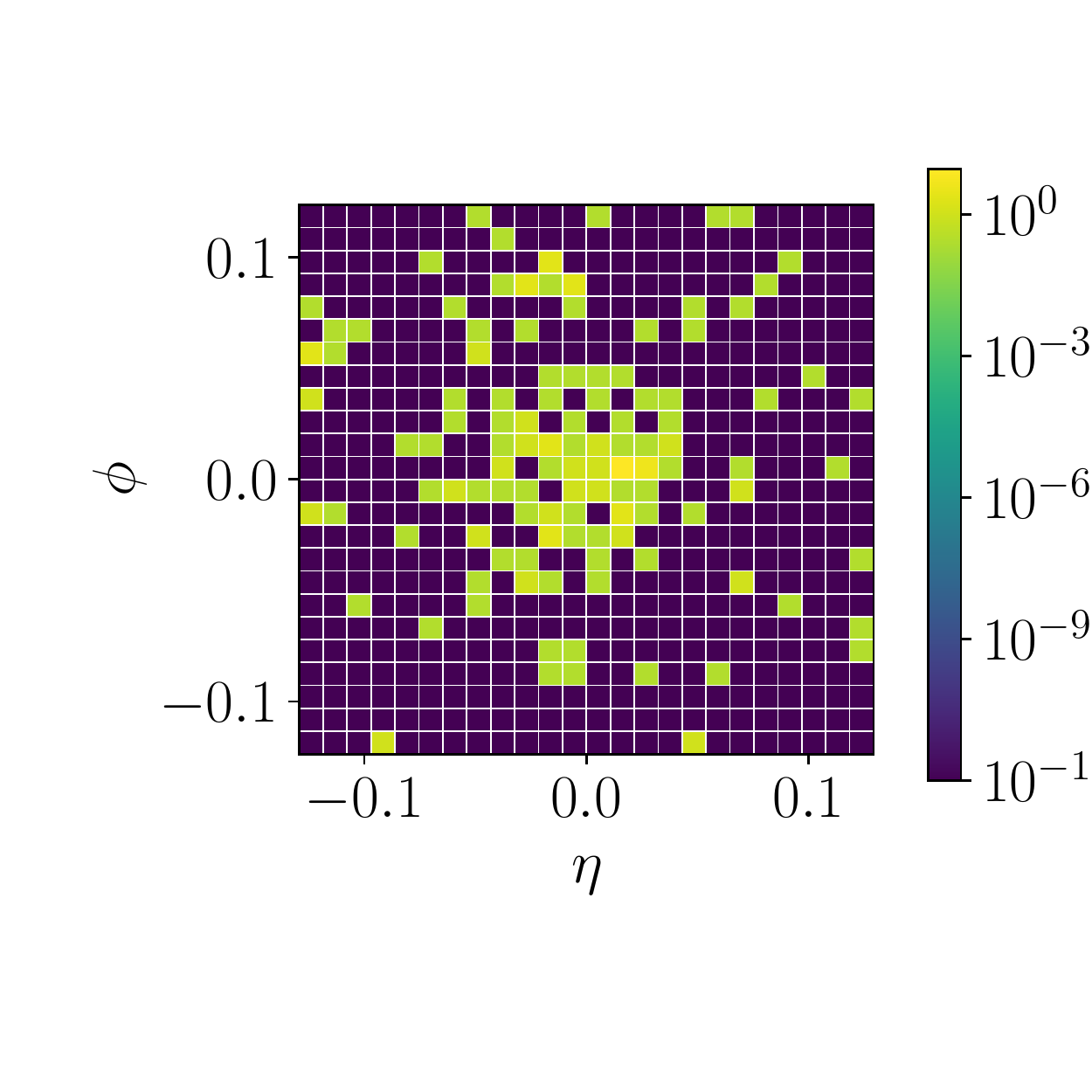}\label{}}\hfill
  \caption{Examples of individual events of the outer layer generated by GAAM}
  \label{gaam_individual_outer}
\end{figure}

\begin{figure}[h!]
  \centering
  \subfloat[Outer layer - (24, 24)]{\includegraphics[width=.5\textwidth]{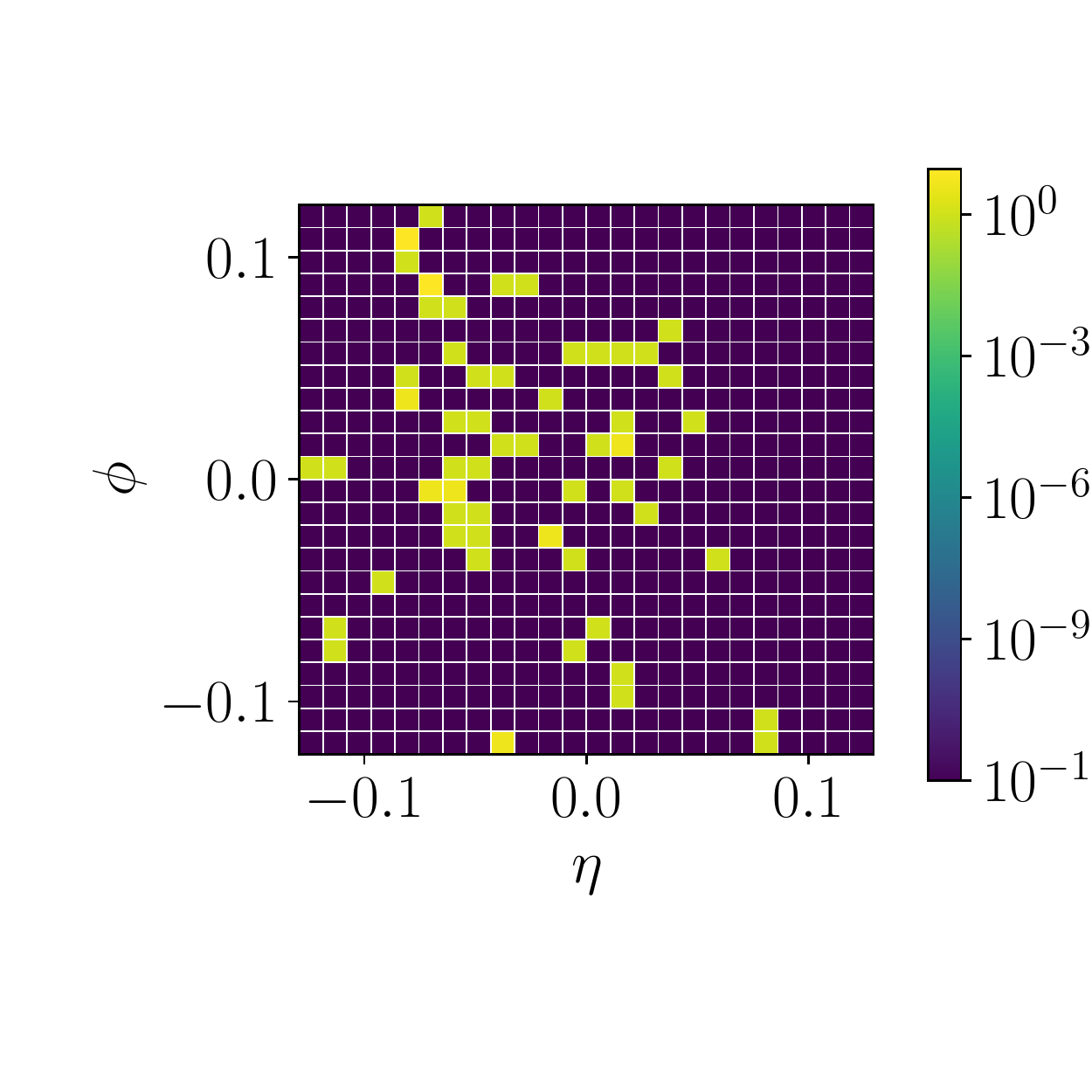}\label{}}\hfill
  \subfloat[Outer layer - (24, 24)]{\includegraphics[width=.5\textwidth]{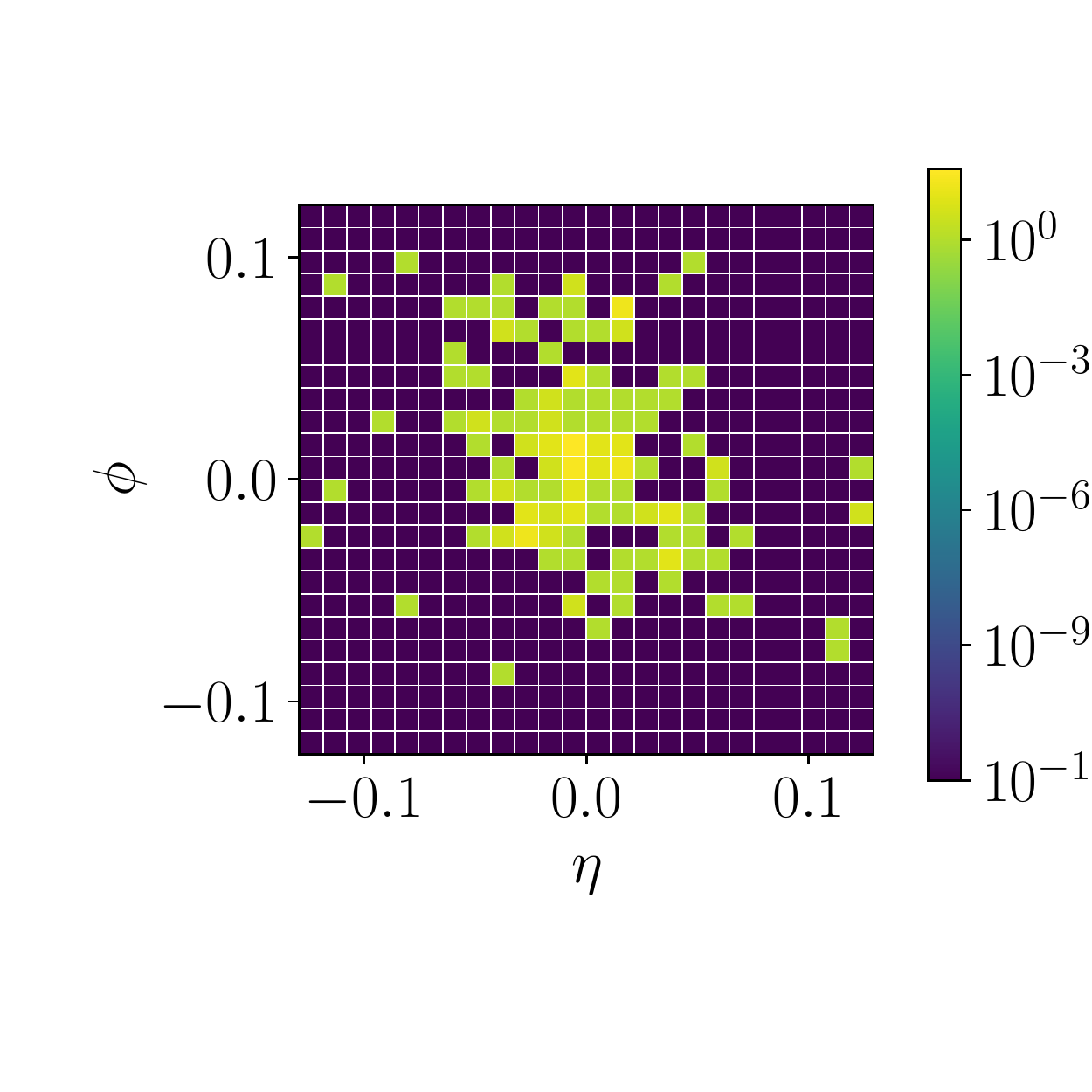}\label{}}\hfill
  \caption{Examples of individual events of the outer layer generated by Geant4}
  \label{true_individual_outer}
\end{figure}

This section shows individual shower images in different geometries and for multiple calorimeter layers. GAAM and Geant4 generated images are shown for the inner layer in Figure~\ref{gaam_individual_inner} and Figure~\ref{true_individual_inner}, middle in Figure~\ref{gaam_individual_middle} and Figure~\ref{true_individual_middle} and the outer layer in Figure~\ref{gaam_individual_outer} and Figure~\ref{true_individual_outer} respectively. 

\end{document}